\documentclass[lettersize,journal]{IEEEtran}
\usepackage[T1]{fontenc}
\usepackage{bm}
\usepackage{makecell}
\usepackage{amssymb,amsmath,amsfonts,amsthm}
\usepackage{mathrsfs}
\usepackage{cite}
\usepackage{array}
\usepackage{tabularx}
\usepackage[table]{xcolor}
\usepackage{multicol, multirow}
\usepackage{color}
\usepackage{mathtools}
\usepackage{subcaption}
\usepackage{mathtools}
\usepackage{tikz,pgf}
\usepackage{verbatim}
\usetikzlibrary{calc,positioning,mindmap,trees,decorations.pathreplacing}
\usepackage{graphicx}
\usepackage{bbm}
\usepackage[utf8]{inputenc}
\usepackage{algorithm}
\usepackage{algorithmic}
\usepackage{booktabs}   
\usepackage{siunitx}    
\usepackage{tikz}
\usetikzlibrary{arrows.meta,calc,positioning}
\newtheorem{theorem}{Theorem}
\newtheorem{lemma}{Lemma}

\newtheorem{remark}{Remark}

\theoremstyle{definition}

\newtheoremstyle{boldplainwithref} 
{\topsep}    
{\topsep}    
{\itshape}   
{}           
{\bfseries}  
{.}          
{ }          
{\thmname{#1}~\thmnumber{#2} \thmnote{#3}}

\theoremstyle{boldplainwithref}

\usepackage[colorinlistoftodos,prependcaption,backgroundcolor=black!5!white,bordercolor=red]{todonotes}

\usepackage{soul}

\usepackage[most]{tcolorbox}

\newtcolorbox{mycomment}[1]{%
	colback=yellow!5,
	colframe=black!60,
	fonttitle=\bfseries,
	title={#1},
	breakable,
}

\begin{document}
	\title{Non-Coherent Over-the-Air Distributed Consensus: Algorithm Design and Power Control Optimization}
	
	\author{ \IEEEauthorblockN{ Yuhang Deng, \textit{Student Member, IEEE}}, \IEEEauthorblockN{Zheng Chen, \textit{Senior Member, IEEE}},\\ and \IEEEauthorblockN{Erik G. Larsson, \textit{Fellow, IEEE}}
		
		\thanks{
			Yuhang Deng, Zheng Chen, Erik G. Larsson are with the Department of Electrical Engineering, Linköping University, 581 83 Linköping, Sweden. (E-mail: \{yuhang.deng, zheng.chen, erik.g.larsson\}@liu.se). 
			(\emph{Corresponding author: Yuhang Deng}.)
			A preliminary version of this paper has been presented at IEEE ICC 2025 \cite{icc25}.\\
			This work was supported in part by Knut and Alice Wallenberg Foundation, ELLIIT, and Swedish Research Council.\\
            This work has been submitted to the IEEE for possible publication. Copyright may be transferred without notice, after which this version may no longer be accessible.
			
		}
	} 
	
	\maketitle  	
	
	\begin{abstract}	
        Non-coherent over-the-air (NCOTA) aggregation offers a scalable and communication-efficient approach to distributed average consensus in wireless networks, without requiring instantaneous channel state information (CSI) or phase alignment. However, persistent aggregation errors pose fundamental challenges in achieving exact average consensus under nominal distributed averaging protocols. In this paper, we develop an \textit{NCOTA consensus framework that accommodates row-stochastic mixing}, without requiring the conventional average-preservation property of doubly stochastic mixing matrices. Building on a distributed optimization formulation of average consensus, we propose a decentralized projected gradient descent with row-stochastic mixing (DPGD-RS) algorithm and establish its mean-square convergence to the exact average of the agents' initial states. We further derive a closed-form upper bound on the mean-square consensus error, characterizing the effects of aggregation uncertainty, network connectivity, and mixing imbalance. Guided by this analysis, we propose a spectral-reshaping power control (SR-PC) scheme that jointly optimizes transmit powers and receive-scaling factors to accelerate convergence by balancing aggregation accuracy and network mixing. Numerical results validate the theoretical convergence guarantees and demonstrate that the proposed power control scheme significantly accelerates convergence compared with full-power transmission and spectral-radius-minimization benchmarks.
	\end{abstract}
	

	\section{Introduction}
	
	\IEEEPARstart{I}{n} multi-agent systems, distributed consensus enables networked agents to agree on a common quantity through local interactions, which is crucial for decentralized coordination tasks such as formation control,  distributed estimation, and sensor fusion \cite{4118472ReZa, 4140748RenWei, 1440896XiaoL, ren2010distributed}. 
    A special case is average consensus, in which the target value is the average of the agents' initial states.
    Classical average consensus protocols are based on distributed linear averaging, where each agent iteratively updates its state by linearly combining its own value with the values received from its neighbors \cite{degroot1974reaching, 1333204Saber, 200465Lin}. 
    The convergence behavior of such systems under perfect communication has been extensively investigated \cite{1333204Saber, 200465Lin, olshevsky2009convergence, 1638541Boyd, 4237149Mehyar} and is jointly determined by the mixing weights and the underlying network topology \cite{200465Lin, olshevsky2009convergence}. 

	In wireless implementations, the idealized assumption of error-free information exchange rarely holds. 
    To account for practical communication impairments, existing studies have investigated average consensus under packet losses and link failures \cite{fagnani2009average, 5499113Shen, 4663899Kar, HUANG20101571}, link-level noise \cite{4663899Kar, HUANG20101571, 06067359XHuang, carli2007average, deng2026exactaverageconsensusnoisy}, and quantization \cite{5325701Kar, 5929538Lavaei}. 
    However, they do not explicitly model inter-agent interference arising from concurrent transmissions in wireless multi-agent systems or the multiple access mechanisms needed to avoid such interference.
    Orthogonal channel access can mitigate inter-agent interference but becomes increasingly inefficient in terms of communication resource usage as the number of active links grows.
	
	\subsection{Coherent vs. Non-Coherent Over-the-Air Aggregation}
	\label{sec:motivation}
	Over-the-air (OTA) computation offers a resource-efficient solution for wireless data aggregation by turning interference into a computation mechanism \cite{10092857sahin, 10155550Chen, 9322286Ozfatura, 9958941Lin, 10154060Nikos}. 
    Specifically, OTA computation leverages the signal superposition in wireless channels to aggregate data from distributed nodes, allowing multiple senders to transmit simultaneously using the same time-frequency resources. 
     This feature is particularly attractive for multi-agent consensus and cooperation with \textit{``many-to-many''} communication topology, where each update is driven by the aggregated information from all neighboring agents in the network. 
     Typical OTA approaches rely on coherent transmission and require accurate instantaneous channel state information (CSI) for amplitude compensation and phase alignment. 
     Such approaches are not directly applicable in systems with more than one receiver, as it is challenging for one sender to align the channel phases at multiple receivers simultaneously, unless additional degrees of freedom are exploited, e.g., through orthogonal time-frequency resource allocation or multi-antenna beamforming as in \cite{9958941Lin}. 
    
    To overcome the challenges of coordinating channel access and acquiring instantaneous CSI for phase alignment in wireless multi-agent systems, non-coherent over-the-air (NCOTA) aggregation has been proposed. 
    The key idea is to encode information in the energy of transmitted signals, allowing the receiver to recover a noisy estimate of the aggregated information from the received signal energy \cite{michelusi2024ota,10607888Yang,11587673Michelusi, 11250607Chen, 11543335Dahl,2026Bozorgasl,Haifeng2026}.
    In a full-duplex setting, NCOTA enables all agents to transmit and receive simultaneously over the same time-frequency resource block, with only coarse time and frequency synchronization required. 
    Specifically, after subtracting the receiver noise energy, the decoded signal at each agent equals, in expectation, the weighted sum of the transmitted signal energies from all agents, with the weights determined by the average channel gains.
    Consequently, the main advantage of NCOTA aggregation for multi-agent consensus systems lies in its highly efficient usage of communication resources and scalability: \textit{no matter how large the network grows, the communication resource consumption for each aggregation remains the same.} 
 		
	\subsection{Closely Related Work}	
	Despite the growing research interest in NCOTA aggregation, only a few recent studies \cite{michelusi2024ota,10607888Yang,11587673Michelusi, 11250607Chen} have explored its application in distributed consensus and optimization. 
    Among these, \cite{10607888Yang, michelusi2024ota} are most closely related to our work. 
    In \cite{10607888Yang}, an energy-based encoding and decoding design was used in combination with the standard linear averaging rule. 
    However, due to the cumulated effects of aggregation noise, the resulting asymptotic consensus value has a random bias, i.e., there is no guarantee to converge to the exact initial average. 
    In a more general setting, NCOTA mechanism was incorporated into a decentralized framework for solving distributed optimization problems \cite{michelusi2024ota}, which is also applicable for the average consensus problem. 
    Nevertheless, all nodes use the same transmit power and the mixing weights are determined purely by the average channel gains.
    As a result, the mixing rate can be far from optimal. 
    The limitations of these two closely related works motivate us to develop a novel processing design for NCOTA average consensus that achieves \textit{convergence to exact average} of initial values with \textit{accelerated convergence rate}. 
	
	\subsection{Contributions}
	The main focus of this work is to develop a wireless average consensus framework based on NCOTA aggregation which is power-efficient and robust to aggregation noise. 
    The main contributions of this work are as follows.
	\begin{itemize}
    \item Building on the NCOTA distributed optimization framework proposed in \cite{michelusi2024ota}, we develop a wireless average consensus protocol that achieves exact convergence to the average of the initial states under persistent aggregation noise.
	
    \item Observing that the noisy consensus recursion is affected by the OTA-induced mixing matrix, we introduce a \textit{transmitter-side power control} scheme combined with \textit{receiver-side scaling} as a ``topology optimization'' technique to accelerate convergence. 
                  
	\item Unlike existing designs \cite{10607888Yang, michelusi2024ota} that rely on doubly stochastic mixing, we adopt \textit{row-stochastic mixing} \cite{MAI201994} to enlarge the power-control design space for faster convergence, while incorporating a local gradient-correction mechanism to ensure convergence to the exact average.
        
    \item We prove that the proposed algorithm achieves mean-square average consensus (MSAC) and derive a closed-form upper bound on the mean-square consensus error, which reveals the key factors governing the convergence behavior and provides theoretical guideline for the power control optimization problem.
	\end{itemize}
	Compared with the conference version \cite{icc25}, we provide substantial extensions in both theoretical analysis and algorithm design. 
    First, the NCOTA average consensus algorithm is generalized to a decentralized projected gradient descent (DPGD) framework, where the OTA-induced mixing matrix is no longer required to be doubly stochastic.  
    Second, we provide a detailed spectral analysis of the mixing matrix and derive a convergence upper bound, considering the joint effects of state-dependent OTA noise, projected update, and row-stochastic mixing.
    Finally, different from the power-control scheme developed in \cite{icc25}, which is retained as a benchmark, we propose a new algorithm named \textit{spectral-reshaping power control (SR-PC)} that directly optimizes the derived convergence upper bound.

\subsection{Notation}
We denote the $i$-th component of vector $\mathbf{x}$ by $[\mathbf{x}]_i$, and the element in the $i$-th row and $j$-th column of matrix $\mathbf{A}$ by $[\mathbf{A}]_{i,j}$. 
The symbols $\mathbf{0}$ and $\mathbf{1}$ denote the all-zero and all-ones vectors, respectively.
We define $\mathbb{N}_0\triangleq\{0,1,2,\ldots\}$ and $\mathbb{R}_+\triangleq(0,\infty)$.
Unless otherwise specified, $\|\cdot\|$ denotes the Euclidean norm for vectors and the spectral norm for matrices. 
For any random vector $\mathbf a$, we define $\|\mathbf a\|_{\mathbb E} \triangleq \sqrt{ \mathbb E [\|\mathbf a\|^2]}$ as its root mean-square norm. 
In particular, $\|\mathbf a\|_{\mathbb E}=\|\mathbf a\|$ if $\mathbf a$ is deterministic.
For any nonempty, compact, and convex set $\mathcal{C} \subseteq \mathbb{R}^d$, where $d\in \mathbb{N}$, we use $\operatorname{bd}(\mathcal C)$ to denote its boundary and define its diameter as $\operatorname{dm}(\mathcal C) \triangleq \max_{\mathbf{x},\, \hat{\mathbf x} \in \mathcal{C}}\| \mathbf{x} - \hat{\mathbf x}\|$. 
For any $\mathbf{z}\in \mathbb{R}^d$, its Euclidean projection onto $\mathcal{C}$ is defined as $\mathcal{P}_{\mathcal{C}}[\mathbf{z}]\triangleq\arg \min _{\mathbf{x} \in \mathcal{C}}\|\mathbf{x}-\mathbf{z}\|$.

\section{Preliminaries}
\label{Preliminaries on Distributed Average Consensus}

\subsection{Distributed Average Consensus}
\label{subsection: Distributed Average Consensus}

Consider a network represented by a complete undirected graph $\mathcal{G} = \left(\mathcal{N},\,\mathcal{E}\right)$, where $\mathcal{N}=\{1,\ldots, N\}$ denotes the set of agents and $\mathcal{E} = \mathcal{N}\times \mathcal{N}$ denotes the set of links.
Each agent holds a real-valued scalar state $x_n[t] \in \mathbb{R}$, where $t \in \mathbb{N}_0$ denotes the iteration index.\footnote{The analysis can be easily generalized to the multi-dimensional case by allowing each state to be a $d$-dimensional vector: $\mathbf{x}_n[t] \in \mathbb{R}^{d}$.}
The goal of distributed consensus is for all agents to agree on a common value through iterative information exchange.
Stacking the agents' states into the vector $\mathbf{x}[t] = [x_1[t],\, x_2[t],\, \ldots,\, x_N[t]]^{\top} \!\in \! \mathbb{R}^{N}$, the system is said to achieve asymptotic consensus if, 
\begin{equation}
	\label{Eq.Consensus}
	\forall \mathbf{x}[0] \in \mathbb{R}^N, \quad \exists a\in \mathbb R: \ \lim _{t \rightarrow \infty} \mathbf{x}[t] = a \cdot \mathbf{1},
\end{equation}
where $a \in \mathbb{R}$ is the consensus value.
A special case of \eqref{Eq.Consensus} is the \textit{average consensus}, where $a = \sum_{n\in\mathcal{N}} x_n[0] / N$ is the average of agents' initial states.  
A classical protocol to achieve distributed consensus is the linear iteration in the form of
\begin{equation}
	\label{Eq.distributed_linear_iteration}
	x_n[t \! + \! 1] = \sum\limits_{m \in \mathcal{N}} w_{n,m} \cdot x_{m}[t],\quad \forall n \in \mathcal{N},
\end{equation}
where $w_{n,m}$ denotes the weight that agent $n$ assigns to the information received from agent $m$.
Let $\mathbf{W}\in \mathbb{R}^{N\times N}$ be a matrix with $ [\mathbf{W}]_{n,m} = w_{n,m}$. Then, \eqref{Eq.distributed_linear_iteration} can be written in a matrix form as
\begin{equation}
	\label{Eq.3}
	\mathbf{x}[t \! + \! 1] = \mathbf{W} \cdot \mathbf{x}[t].
\end{equation}
For the distributed linear iteration in \eqref{Eq.3}, it achieves average consensus if $\mathbf{W} \mathbf{1}=\mathbf{1}$, $\mathbf{1}^{\top} \mathbf{W}=\mathbf{1}^{\top}$, and $\rho(\mathbf{W}-\frac{\mathbf{1} \mathbf{1}^{\top}}{N})<1$, where $\rho(\cdot)$ denotes the spectral radius.
In addition, a smaller $\rho(\mathbf{W}-\frac{\mathbf{1} \mathbf{1}^{\top}}{N})$ leads to faster asymptotic convergence of \eqref{Eq.3}.
These conditions are recovered by setting $\mathbf{u} = \mathbf 1$ in Lemma \ref{Lemma.1}, which specializes the convergence result in \cite[Theorem 2]{200465Lin}. 

\begin{lemma}[A special case of {\cite[Theorem 2]{200465Lin}}]
	\label{Lemma.1}
	Let $\mathbf W\in\mathbb R^{N\times N}$ and $\mathbf u\in\mathbb R^{N}$ with $\mathbf u^{\top}\mathbf 1\neq 0$. If the following three conditions hold: \textup{(S1)}~$\mathbf{W} \cdot \mathbf{1} =\mathbf{1}$, \textup{(S2)}~$\mathbf{u}^{\top} \!\cdot\! \mathbf{W}=\mathbf{u}^{\top}$, \textup{(S3)}~$\rho\left(\mathbf{W} - \mathbf{1} \mathbf{u}^{\top} /  \mathbf{u}^{\top} \mathbf{1}\right) <1$, we have
	\begin{equation}
		\lim_{t\to\infty}\mathbf W^{t}
		=\frac{\mathbf{1}\, \mathbf{u}^{\top}}{\mathbf{u}^{\top} \mathbf{1}}.
	\end{equation}
	Consequently, the iteration $\mathbf{x}[t \!+\!1]=\mathbf {W}\,\mathbf{x}[t]$ converges to consensus value when $t \rightarrow \infty$:
	\begin{equation}
		\lim\limits_{t \rightarrow \infty} \mathbf{x}[t] = \lim\limits_{t \rightarrow \infty} \mathbf{W}^t \cdot \mathbf{x}[0] = \frac{\mathbf{1}\, \mathbf{u}^{\top}}{\mathbf{u}^{\top} \mathbf{1}} \cdot \mathbf{x}[0] =a \cdot \mathbf{1},
	\end{equation}
	where the consensus value is $a=\mathbf{u}^{\top}\mathbf{x}[0] /  \mathbf{u}^{\top} \mathbf{1}$.
\end{lemma}
Under conditions \textup{(S1)--(S3)}, the eigenvalue $1$ of $\mathbf W$ is simple and strictly dominant, and $\mathbf{u}$ is the corresponding left eigenvector that is unique up to scaling. The limiting consensus value is determined by the normalized left eigenvector.

Note that another way of writing the consensus averaging protocol in \eqref{Eq.distributed_linear_iteration} is 
\begin{equation}
    \label{Eq.pairwise}
    x_n[t \! + \! 1] = x_n[t] + \!\! \sum\limits_{m \in \mathcal{N}_n} \! w_{n,m} \left(x_{m}[t]-x_{n}[t]\right),\  \forall n \in \mathcal{N},
\end{equation}
where $\mathcal{N}_n=\{m\in\mathcal{N}:(n,m)\in \mathcal{E}\}$ denotes the neighbor set of agent $n$.
With this formulation, we have $w_{n,n}=1-\sum\nolimits_{m \in \mathcal{N}_n} w_{n,m}$, which naturally ensures $\mathbf{W} \mathbf{1} =\mathbf{1}$. If \textup{(S2)} and \textup{(S3)} are satisfied with $\mathbf u = \mathbf 1$, \eqref{Eq.pairwise} achieves average consensus.

\subsection{Average Consensus as a Distributed Optimization Problem}
Let $\mathbf{x} = [x_1,\,\ldots,\, x_N]^{\top}$ denote the collection of local optimization variables held by the agents.
The average consensus problem can be equivalently reformulated as the following distributed optimization problem: 
\begin{subequations}
\label{Eq.DGD_problem}
\begin{align}
	&\min_{\mathbf{x}} \   f(\mathbf{x}) \coloneqq  \sum\limits_{n \in \mathcal{N}} \, \frac{1}{2}\left(x_n - x_n[0]\right)^2, \\
	&\text { s.t. } 	\quad  x_1 = \cdots = x_N.
\end{align}
\end{subequations}
We define $f_{n}(x_n) = \frac{1}{2}\left(x_n - x_n[0]\right)^2 $ as the quadratic local objective function of the $n$-th agent, of which the Hessian satisfies $\nabla^2 f_n(x_n)=1$ for all $x_n$. 
Thus, each $f_n(x_n)$ is $1$-strongly convex and $1$-smooth.
Under the consensus constraint, the global objective $f(\mathbf{x})$ has the unique minimizer $\mathbf{x}^* = x^*\, \mathbf 1$, where $ x^* = \sum_{n \in \mathcal{N}}x_n[0]/N$ is the average of initial states.
Therefore, any distributed optimization algorithm that solves \eqref{Eq.DGD_problem} can lead to average consensus.
In the following, we introduce two distributed optimization algorithms for solving \eqref{Eq.DGD_problem} under different assumptions on the mixing matrix.

\subsubsection{Decentralized gradient descent (DGD) with doubly stochastic mixing \cite{4749425Nedic}}
The general form of DGD is 
\begin{equation}
\label{Eq.DGD}
\mathbf{x}[t \! + \! 1]= \mathbf{W} \cdot \mathbf{x}[t] - \eta[t] \cdot \mathbf{d}[t],
\end{equation}
where $\mathbf{d}[t]= [\nabla f_{1}(x_1[t]), \, \ldots,\, \nabla f_{N}(x_N[t])]^{\top}$ collects the local gradients of the agents and $\eta[t]$ denotes the stepsize. 
The standard DGD algorithm commonly adopts a doubly stochastic mixing matrix $\mathbf W$, which satisfies the conditions \textup{(S1)} and \textup{(S2)} in Lemma \ref{Lemma.1} with $\mathbf u = \mathbf 1$.\footnote{In fact, non-negativity of $\mathbf W$ is not required for convergence, as shown in our previous work \cite{icc25}.}
For \eqref{Eq.DGD} to converge to $\mathbf{x}^* = x^* \mathbf 1$, $\mathbf{W}$ is also required to satisfy \textup{(S3)}. 
In addition, the stepsize sequence is required to satisfy the standard diminishing-stepsize condition \cite{lobel2011distributed}: \textup{(S4)} $\{\eta[t]\}_{t=0}^{\infty}$ is a positive non-increasing sequence such that
\begin{equation}
\sum_{t=0}^{\infty}\eta[t]=\infty ~~\text{and} ~~\sum_{t=0}^{\infty}\eta^2[t]<\infty.
\end{equation}
Under these conditions, the asymptotic convergence behavior of \eqref{Eq.DGD} depends on both the spectral radius $\rho(\mathbf{W} - \frac{\mathbf{1} \mathbf{1}^\top}{N})$, i.e., the second largest modulus eigenvalue of $\mathbf{W}$, and the stepsize sequence $\{\eta[t]\}$.

\subsubsection{DGD with row-stochastic mixing and gradient correction}
Relaxing the doubly stochastic mixing requirement is desirable, for example, when the communication graph is directed\footnote{The construction of a doubly stochastic matrix for arbitrary directed graphs is generally difficult, and often infeasible.} or when enlarging the feasible design space of $\mathbf{W}$ can potentially lead to faster convergence.
Here we introduce DGD with row-stochastic mixing (DGD-RS) \cite{MAI201994}.

Recall that $\mathbf{u} = [u_1,\, \ldots,\, u_N]^{\top}$ is the left eigenvector of $\mathbf W$ associated with eigenvalue $1$, i.e., $\mathbf{u}^{\top} \mathbf{W}=\mathbf{u}^{\top}$. The general form of DGD-RS is 
\begin{equation}
\label{Eq.DGD-RS}
\mathbf{x}[t \! + \! 1]= \mathbf{W} \cdot \mathbf{x}[t] - \eta[t] \cdot h_{\mathbf{u}} \cdot \mathbf{d}_{\mathbf{u}}[t],
\end{equation}
where $\mathbf{d}_{\mathbf{u}}[t]= [\nabla f_{1}(x_1[t])/u_1,\, \ldots,\, \nabla f_{N}(x_N[t])/u_N]^{\top}$ is the left-eigenvector-scaled local gradient vector, which will be referred to as the \textit{corrected gradient}. 
When $\mathbf{W}$ is non-negative and irreducible, the Perron–Frobenius theorem guarantees that $u_n > 0$ for all $n\in\mathcal N$. Hence, the corrected gradient is well defined with positive weights.
Here, we introduce an additional normalization factor $h_{\mathbf u} \triangleq N / \sum_{n \in \mathcal{N}} \left(1/u_n\right)$, which is the harmonic mean of $\mathbf u$.
We assume that $\mathbf{W}$ satisfies conditions \textup{(S1)--(S3)} in Lemma~\ref{Lemma.1}, and that the stepsize sequence $\{\eta[t]\}$ satisfies \textup{(S4)}. 

For each agent $n \in \mathcal{N}$, the local objective is $f_n(x_n[t])=\frac{1}{2}(x_n[t] - x_n[0])^2$, which yields $\nabla f_n(x_n[t]) = x_n[t] - x_n[0]$. Therefore, the per-agent update rule can be expressed as the following linear recursion:
\begin{equation}
	\label{Eq.DGD-RS-agentwise}
	x_n[t+1] = \!\! \sum_{m \in \mathcal{N}} w_{n,m} \,x_m[t] - \frac{\eta[t] h_{\mathbf{u}}}{u_n}\left(x_n[t] - x_n[0]\right).
\end{equation}
In general, $u_n$ and $h_{\mathbf{u}}$ are not locally available and may require online estimation. In our OTA implementation, however, they can be determined in advance for faster convergence through the optimized power-control and receive-scaling design.

The effect of row-stochastic mixing and corrected gradient can be interpreted as follows. Condition \textup{(S3)} gives $\rho(\mathbf{W} - \frac{\mathbf{1} \mathbf{u}^{\top}} {\mathbf{u}^{\top} \mathbf{1}})<1$, which ensures that the repeated application of $\mathbf{W}$ drives every vector in the disagreement subspace $\operatorname{ker}\{\mathbf{u}^{\top}\}$ to zero.
On the other hand, left-multiplying \eqref{Eq.DGD-RS} by $\mathbf{u}^{\top}$ yields
\begin{align}
\label{Eq.sum_DGD-RS}
\mathbf{u}^{\top}\mathbf{x}[t \! + \! 1] & = \mathbf{u}^{\top}\mathbf{W} \cdot \mathbf{x}[t] - \eta[t] \, h_{\mathbf{u}} \cdot \mathbf{u}^{\top} \mathbf{d}_{\mathbf{u}}[t] \\
& = \mathbf{u}^{\top} \mathbf{x}[t] - \eta[t] \, h_{\mathbf{u}} \sum\nolimits_{n\in\mathcal{N}}\nabla f_n(x_n[t]). \nonumber
\end{align}
Thus, the left-eigenvector scaling recovers the sum of the local gradients in the aggregate dynamics.
As the disagreement among the agents vanishes, the local states approach a common value, and the aggregate gradient term in \eqref{Eq.sum_DGD-RS} approaches the centralized gradient $\sum_{n\in\mathcal{N}}\nabla f_n(x)$. 
Thus, the aggregate dynamics asymptotically reduce to the centralized gradient descent iteration for solving \eqref{Eq.DGD_problem}, which explains how the left-eigenvector scaling compensates for the imbalance introduced by row-stochastic mixing \cite{MAI201994}.


\section{Distributed Consensus via Non-Coherent Over-the-Air Aggregation}
\label{Non-Coherent Over-the-Air Aggregation}

Implementing distributed consensus protocols in wireless networks is challenging because of the ``many-to-many'' communication topology: all nodes act as senders and receivers in every iteration. 
Concurrent transmissions over shared wireless medium introduce interference that deteriorates communication quality. Traditionally, interference can be mitigated through orthogonal resource allocation and scheduling, as shown in Fig. \ref{fig:wireless_access_mechanisms}(a) and \ref{fig:wireless_access_mechanisms}(b). 
In particular, coherent OTA aggregation (Fig. \ref{fig:wireless_access_mechanisms}(b)) for distributed optimization has been considered in \cite{9322286Ozfatura, 10154060Nikos}.
However, they  requires assigning orthogonal resources to different receivers as a way to parallelize the computing process.
In contrast, NCOTA aggregation (Fig. \ref{fig:wireless_access_mechanisms}(c)) offers significant advantages in terms of resource efficiency and scalability, as it enables simultaneous transmissions from all nodes without separating them in frequency or time.

\definecolor{slot1}{RGB}{214,39,40}   
\definecolor{slot2}{RGB}{31,119,180}  
\definecolor{slot3}{RGB}{44,160,44}   

\tikzset{ 
    nodeStyle/.style={ circle, draw=black, fill=white, minimum size=8mm, inner sep=0pt, font=\small },
    arrowStyle/.style={ -{Latex[length=2.2mm,width=1.7mm]}, line width=1.0pt },
    captionStyle/.style={ font=\small, align=center, text width=5.0cm}
}

\newcommand{\ThreeNodeNodes}{
\node[nodeStyle] (v1) at ( 0,  1.4) {$1$};
\node[nodeStyle] (v2) at (-1.4, -0.6) {$2$};
\node[nodeStyle] (v3) at ( 1.4, -0.6) {$3$};
}

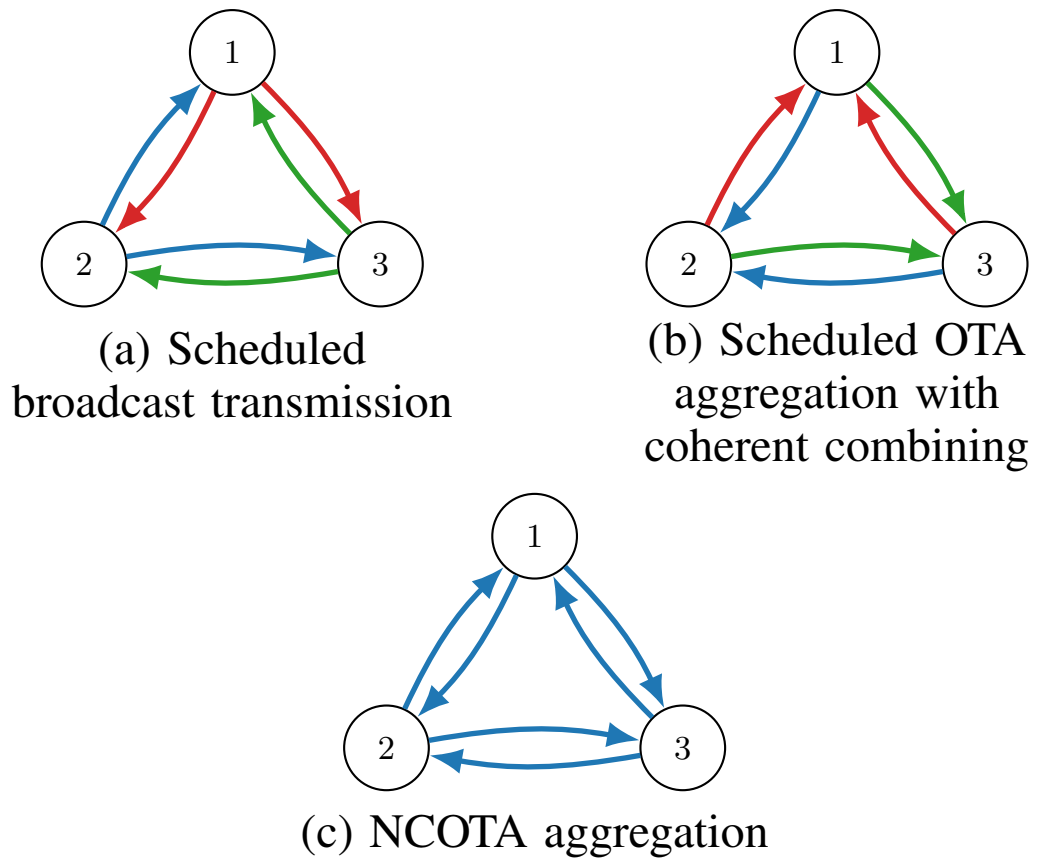
\begin{figure}[t]
	\centering
	\begin{tikzpicture}[x=1cm,y=1cm]
		
		\begin{scope}[shift={(-2,0)}]
			\begin{scope}[scale=0.7, transform shape]
				\ThreeNodeNodes
				
				\draw[arrowStyle,slot1] (v1) to[bend left=10] (v2);
				\draw[arrowStyle,slot1] (v1) to[bend left=10] (v3);
				
				\draw[arrowStyle,slot2] (v2) to[bend left=10] (v1);
				\draw[arrowStyle,slot2] (v2) to[bend left=10] (v3);
				
				\draw[arrowStyle,slot3] (v3) to[bend left=10] (v1);
				\draw[arrowStyle,slot3] (v3) to[bend left=10] (v2);
			\end{scope}
			
			\node[font=\small, align=center, text width=3.0cm] at (0,-1.15)
			{(a) Scheduled\\broadcast transmission};
		\end{scope}
		
		\begin{scope}[shift={(2,0)}]
			\begin{scope}[scale=0.7, transform shape]
				\ThreeNodeNodes
				
				\draw[arrowStyle,slot1] (v2) to[bend left=10] (v1);
				\draw[arrowStyle,slot1] (v3) to[bend left=10] (v1);
				
				\draw[arrowStyle,slot2] (v1) to[bend left=10] (v2);
				\draw[arrowStyle,slot2] (v3) to[bend left=10] (v2);
				
				\draw[arrowStyle,slot3] (v1) to[bend left=10] (v3);
				\draw[arrowStyle,slot3] (v2) to[bend left=10] (v3);
			\end{scope}
			
			\node[font=\small, align=center, text width=3.0cm] at (0,-1.3)
			{(b) Scheduled OTA \\ aggregation with \\ coherent combining};
		\end{scope}
		
		\begin{scope}[shift={(0,-3.2)}]
			\begin{scope}[scale=0.7, transform shape]
				\ThreeNodeNodes
				
				\draw[arrowStyle,slot2] (v1) to[bend left=10] (v2);
				\draw[arrowStyle,slot2] (v2) to[bend left=10] (v1);
				
				\draw[arrowStyle,slot2] (v1) to[bend left=10] (v3);
				\draw[arrowStyle,slot2] (v3) to[bend left=10] (v1);
				
				\draw[arrowStyle,slot2] (v2) to[bend left=10] (v3);
				\draw[arrowStyle,slot2] (v3) to[bend left=10] (v2);
			\end{scope}
			
			\node[font=\small, align=center, text width=6.0cm] at (0,-1)
			{(c) NCOTA aggregation};
		\end{scope}
		
	\end{tikzpicture}
	\caption{Comparison of three wireless access mechanisms. Different colors represent orthogonal resource blocks.}
	\label{fig:wireless_access_mechanisms}
\end{figure}

In this section, we review the basic framework of decentralized projected gradient descent (DPGD) with NCOTA aggregation, as originally proposed in \cite{michelusi2024ota}, and derive the \textit{noisy linear recursion} for distributed average consensus. Furthermore, we introduce a novel transmit-power control and receive-scaling design, which reshapes the OTA-induced mixing weights, thereby altering the effective network connectivity and its spectral properties. It also determines the left-eigenvector scaling used for gradient correction. These relationships provide the basis for accelerating the convergence to average consensus.

\subsection{NCOTA Distributed Consensus via Energy-Based Encoding}
We model the wireless multi-agent network as a complete graph with time-varying link weights, where each link weight reflects the instantaneous quality of the corresponding wireless channel.
All agents are assumed to have full-duplex capability with perfect self-interference cancellation, so that they can transmit and receive simultaneously without being affected by their own transmission.\footnote{The full-duplex capability is assumed for simplicity. The NCOTA framework can accommodate half-duplex operation through randomized transmit/receive selection, as demonstrated in \cite{michelusi2024ota}.}
The complex channel coefficient from agent $m$ to $n$ at time index (or iteration) $t$ is denoted by
\begin{equation}
g_{n,m}[t] = h_{n,m}[t] \sqrt{\beta_{n,m}} \in\mathbb{C},
\end{equation}  
where $\beta_{n,m} \!\in\! \mathbb{R}_+$ is the large-scale fading coefficient and $h_{n,m}[t]  \sim  \mathcal{CN}(0, 1)$ is the small-scale fading. We assume that $h_{n,m}[t]$ is independent over time and across links, and that the large-scale fading coefficients are reciprocal: $\beta_{n,m}= \beta_{m,n}$.

\begin{figure*}[h!]
\centering
\includegraphics[width=0.85\linewidth]{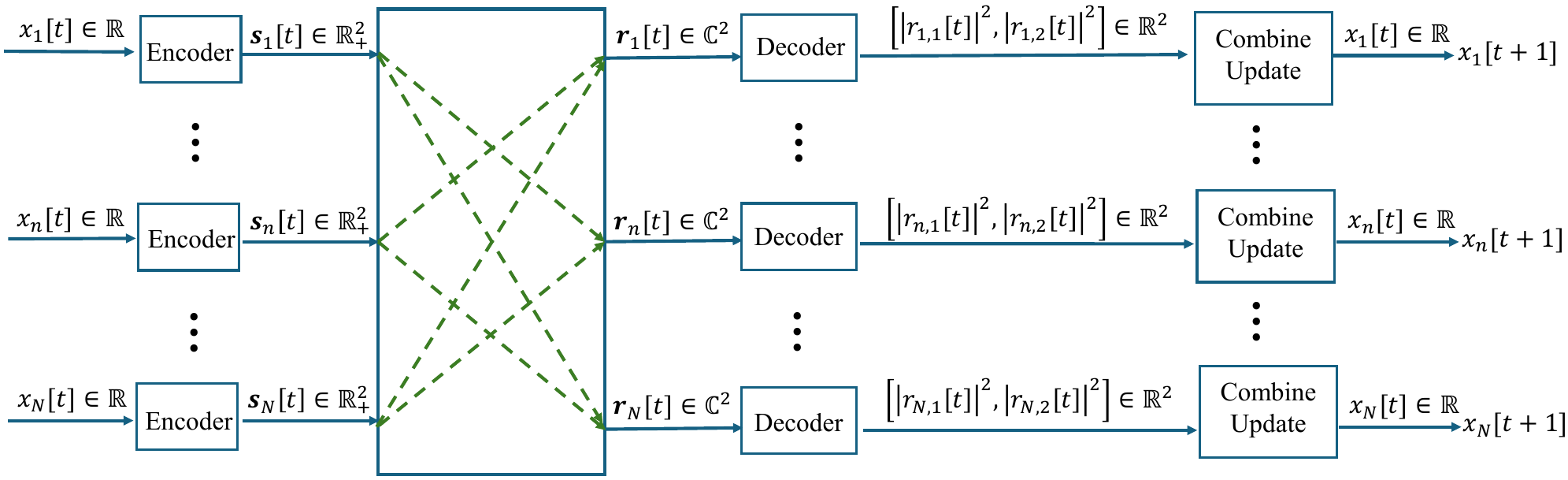}
\caption{Computation process for distributed average consensus with NCOTA aggregation over wireless fading channels.}
\vspace{-1mm}
\label{Fig.1:System}
\end{figure*}

Fig. \ref{Fig.1:System} illustrates the processing chain of NCOTA average consensus, with the main steps summarized as follows:
\begin{itemize}
\item Prior to transmission, each agent encodes its state (source data) into a two-dimensional non-negative channel symbol using amplitude modulation. 
\item All agents simultaneously transmit their encoded two-dimensional channel symbols over two resource blocks that experience the same channel fading coefficient.
\item Upon receiving the superimposed signals, each agent performs energy-based decoding and combines the decoded information with its current state for the local update.
\end{itemize}
Since this design uses symbol energy to convey information, it does not require instantaneous CSI or phase alignment, but only coarse time-frequency synchronization and the receiver noise power $\sigma^2$. 

We next describe each processing step in detail.

\subsubsection{Source Precoding and Modulation}
\label{subsection: Pre-coding and signal modulation}

The precoding block is a special case of the encoding design proposed in \cite{michelusi2024ota}. We use a codebook $\mathcal{Z} = \{z_1,z_2\}=\{-x_{\max},x_{\max}\}$ with two codewords for precoding, which is known to all agents.  
For any state $x_n[t]\in[-x_{\max},x_{\max}]$, we define
\begin{equation}
p_{n,1}[t]=\frac{x_{\max}-x_n[t]}{2x_{\max}},\qquad
p_{n,2}[t]=\frac{x_{\max}+x_n[t]}{2x_{\max}}.
\end{equation}
It is readily verified that  $p_{n,1}[t]\geq 0$, $p_{n,2}[t]\geq 0$, and $p_{n,1}[t]+p_{n,2}[t]=1$. Moreover, $x_n[t]$ can be represented as a convex combination of the codewords in $\mathcal{Z}$ as
\begin{equation}
\label{Eq.coding}
x_{n}[t] = p_{n,1}[t]\,z_1+p_{n,2}[t]\,z_2, \quad \forall \, n \in \mathcal{N},\ t \in \mathbb{N}_0.
\end{equation} 
Thus, the original (possibly negative) state value $x_n[t]$ is encoded into a two-dimensional nonnegative vector $ \mathbf{p}_{n}[t] = [p_{n,1}[t],\, p_{n,2}[t]]^{\top}$. 
Let $\bm{\alpha} = [\alpha_{1},\, \alpha_{2},\, \ldots,\, \alpha_{N}]^{\top}$ denote the transmit power control vector, where $\alpha_{n} > 0$ for all $n\in\mathcal{N}$. 
Given the power control factor $\alpha_n$, the encoded channel symbol from agent $n$ at $t$-th iteration is
\begin{equation}
\label{Eq.transmitted_symbol}
\mathbf{s}_n[t] \! = \! \left[s_{n,1}[t], s_{n,2}[t]\right]^{\top} \!\!=\!\! \big[\sqrt{\alpha_{n} p_{n,1}[t]}, \sqrt{\alpha_{n} p_{n,2}[t]}\big]^{\top}.
\end{equation}
The entries of $\mathbf{s}_n[t]$ are transmitted as signal amplitudes using amplitude modulation. 
This channel symbol represents the transmit-power allocation across the two codewords of $\mathcal{Z}$, with total power $\|\mathbf{s}_n[t]\|^2=\alpha_n$.
Introducing transmit power control is one of the key differences from the original design in \cite{michelusi2024ota}.

\subsubsection{Signal Superposition and Energy-Based Decoding} 
While transmitting the encoded channel symbols, each agent simultaneously receives the superposed signals from all other agents. With full duplex capability and perfect self-interference cancellation, the received superimposed signal is a complex-valued two-dimensional vector
\begin{equation}
\label{Eq.received_symbol}	
\mathbf{r}_n[t]=[r_{n,1}[t],\, r_{n,2}[t]]^{\top} = \sum\limits_{\substack{m=1 \\ m\neq n}}^{N} g_{n,m}[t] \cdot \mathbf{s}_{m}[t] + \mathbf{n}_n[t],
\end{equation}
where $\mathbf{n}_n[t] \!\sim\! \mathcal{CN}(\mathbf{0}, \sigma^2 \mathbf{I})$ is the additive noise, independent across agents and iterations and from the channel coefficients. 

Using the received two-dimensional symbol $\mathbf{r}_n[t]$, agent $n$ obtains the squared magnitude of each element in $\mathbf{r}_n[t]$.
For the $i$-th element of $\mathbf{r}_n[t]$ with $i \in \{1,\,2\}$, we have
\begin{align}
\label{Eq.received_symbol^2}	
&|r_{n,i}[t]|^{2} \nonumber \\
=& \sum\limits_{\substack{m=1 \\ m\neq n}}^{N} \beta_{n,m} \, \lvert h_{n,m}[t]\rvert^{2} \, \alpha_m \, p_{m,i}[t] + [\mathbf{N}_n[t]]_i \\ 
+& \sum\limits_{\substack{m=1 \\ m\neq n}}^{N} \sum\limits_{\substack{k=1 \\ k\neq m,n}}^{N}  \!\!\! \text{Re}\! \left( h_{n,m}^{}[t] h_{n,k}^{*}[t]\right)\!\! \sqrt{ \! \beta_{n,m} \alpha_m p_{m,i}[t] \, \beta_{n,k} \alpha_k p_{k,i}[t]}, \nonumber
\end{align}
where $\! [\mathbf{N}_n[t]]_i \!=\!\! 2 \text{Re}\big(\!\sum_{m=1,  m\neq n}^{N} \! g_{n,m}[t] \! \sqrt{\! \alpha_m p_{m,i}[t]}[\mathbf{n}_n^{*}[t]]_i\big) +  |[\mathbf{n}_n[t]]_i|^{2}$ is the $i$-th element of $\mathbf{N}_n[t] \in \mathbb{R}^{2}$. 
Since $\{g_{n,m}[t]\}$ and $\{\mathbf{n}_n[t]\}$ are independent, $\mathbb{E}[\mathbf{N}_n[t]] \!=\! \sigma^2 \mathbf{1}$.
Using the computed energies $\{|r_{n,i}[t]|^{2}\}_{i=1}^{2}$, each agent applies the following decoding rule: 
\begin{equation}
\label{Eq.update_vector1}
c_n[t] \!=\! \sum\nolimits_{i=1}^{2} (	|r_{n,i}[t]|^2 \!-\! \sigma^2 ) \left(z_i \!-\! x_n[t]\right) \in \mathbb{R}, \  \forall n \in \mathcal{N}.
\end{equation}
\begin{remark}
\label{remark: estimate error for sigma}
For simplicity, we assume that the receiver noise power $\sigma^2$ is accurately known at each agent. In practical systems, thermal noise power typically remains relatively stable over time and can be estimated once for all iterations. Moreover, the proposed approach remains applicable when the noise power is estimated imperfectly, provided that the estimate is conditionally unbiased.
\end{remark}

Substituting the energy-based decoding expression \eqref{Eq.received_symbol^2} into \eqref{Eq.update_vector1}, we decompose $c_n[t]$ as
\begin{equation}
c_n[t]=c_{1,n}[t]+c_{2,n}[t]+c_{3,n}[t],
\end{equation}
where
\begin{align}
c_{1,n}[t]
&= \! \sum_{i=1}^{2} \sum\limits_{\substack{m=1 \\ m\neq n}}^{N} \lvert h_{n,m}[t]\rvert^{2} \beta_{n,m} \alpha_m p_{m,i}[t] \left(z_{i} \!-\! x_n[t] \right), \label{Eq.c1n} \\
c_{2,n}[t]
&= \! \sum\limits_{\substack{m=1 \\ m\neq n}}^{N} \sum\limits_{\substack{k=1 \\ k\neq m,n}}^{N}  \text{Re}\left(h_{n,m}^{}[t] h_{n,k}^{*}[t]\right) \sqrt{\beta_{n,m}\beta_{n,k}\alpha_m \alpha_k}  \notag \\ &\quad \Big(\sum\nolimits_{i=1}^{2}\sqrt{p_{m,i}[t] p_{k,i}[t]}\left(z_i - x_{n}[t]\right)\Big), \label{Eq.c2n} \\
c_{3,n}[t]
&= \!\sum\nolimits_{i=1}^{2}([\mathbf{N}_n[t]]_i -\sigma^2) \left(z_i - x_n[t]\right). \label{Eq.c3n}
\end{align}
Here, $c_{1,n}[t]$ is the desired energy term, $c_{2,n}[t]$ is the non-coherent interference term caused by concurrent transmissions, and $c_{3,n}[t]$ is the noise-related term. 
Due to the randomness of the small-scale fading and receiver noise, $\{c_n[t] \, \vert \, t \in \mathbb{N}_0\}$ is a stochastic process. 
Let $\mathcal F_t$ denote the $\sigma$-algebra generated by all states, channel realizations, and noise samples up to the iteration $t-1$. 
Taking the expectation over the randomness in the $t$-th iteration, we can characterize the conditional expectation of $c_n[t]$ given $\mathcal F_t$.
\begin{itemize}
\item Since $h_{m,n}[t] \!\sim\! \mathcal{CN}(0,1)$ and is independent over time, $\mathbb{E}[\lvert h_{m,n}[t]\rvert^{2} \vert \mathcal{F}_t] \!=\! 1$ holds.
Substituting \eqref{Eq.coding} in \eqref{Eq.c1n} yields
\begin{equation}
\label{Eq.ExpectedC1N}	
\mathbb{E}[c_{1,n}[t] \,\vert\, \mathcal{F}_t] \!=\! \sum\nolimits_{\substack{m=1 \\ m\neq n}}^{N} \beta_{n,m} \alpha_m \left( x_{m}[t] - x_n[t] \right),
\end{equation}
which is a weighted sum of the differences between the $n$-th agent's state and those of its neighbors.

\item  Given that $h_{n,m}[t]$ is zero-mean and i.i.d. for all $n,m \in \mathcal{N}$ and $n\neq m$, we have $\mathbb{E}[c_{2,n}[t]\vert \mathcal{F}_t] = 0$.

\item  Since $\mathbb{E}[\mathbf{N}_n[t]\vert \mathcal{F}_t] = \sigma^2 \,\mathbf{1}$, we have $\mathbb{E}[c_{3,n}[t] \vert \mathcal{F}_t] = 0$. 	
\end{itemize}

Finally, we have
\begin{equation}
\label{Eq.ExpectedCN}	
\mathbb{E}[c_n[t] \vert \mathcal{F}_t] = \sum\nolimits_{\substack{m=1 \\ m\neq n}}^{N} \beta_{n,m} \alpha_{m} \left(x_m[t]-x_n[t]\right).   
\end{equation}
\begin{remark}
The energy-based decoding rule \eqref{Eq.update_vector1} yields a conditionally unbiased estimator of a weighted sum of pairwise state differences, with the weights determined by the large-scale fading  and the transmit-power control coefficients. This formulation resembles the classical linear consensus averaging rule in \eqref{Eq.pairwise}, but in a noisy setting.
\end{remark}

\subsubsection{Information Combining and State Updating}
In the final step, each agent combines the decoded information with its local state value. 
We define $\bm{\gamma} = [\gamma_{1},\, \gamma_{2},\, \ldots,\, \gamma_{N}]^{\top}$ as the receive scaling vector with $\gamma_{n} > 0$ for all $n \in \mathcal {N}$. After each agent $n$ computes $c_n[t]$ from \eqref{Eq.update_vector1}, it updates the state value  $x_n[t+1]$ according to the following iteration rule:
\begin{equation}
\label{Eq.UpdatedRule1}	
x_{n}[t + 1] \!=\!\mathcal{P}_{\Omega_n}\!\Big[\!\!\!\underbrace{x_n[t]}_{\text{local state}} \!+\!\!\!\! \underbrace{\zeta[t] \gamma_n c_n[t]}_{\text{ innovation from neighbors}} \!\!\!- \! \underbrace{\eta[t] h_{\bm\alpha,\bm\gamma}\, \frac{\gamma_n d_n[t]}{\alpha_n}}_{\text{local-gradient correction}}\Big].
\end{equation}
There are two stepsize sequences in \eqref{Eq.UpdatedRule1}: $\zeta[t]$ is the consensus stepsize applied to $c_n[t]$, and $\eta[t]$ is the gradient stepsize applied to $d_n[t]$. Both stepsizes satisfy \textup{(S4)}.
The feasible sets $\Omega_n$ are identical for all agents and given by the convex hull of $\mathcal{Z}$: $\Omega_n = \operatorname{conv}\left(\mathcal{Z}\right) = [-x_{\max}, x_{\max}]$, $\forall n \in \mathcal{N}$. 
It follows that $\operatorname{bd}(\Omega_n) = \{-x_{\max},x_{\max}\}$ and $\operatorname{dm}(\Omega_n) = 2x_{\max}$.
This projection ensures that the updated state remains within the codebook's range, i.e., $|x_n[t+1]| \leq x_{\max}$. Finally, $d_n[t]$ represents the local gradient of agent $n$, which can be directly computed as 
\begin{equation}
d_n[t] \!=\! \nabla f_{n}(x_n[t]) \!=\! x_{n}[t] \!-\! x_{n}[0].
\end{equation}
The scalar $h_{\bm\alpha,\bm\gamma} = N/(\sum_{n=1}^{N} \gamma_n/\alpha_n)$ normalizes the weights of the local-gradient correction.
This gradient correction follows the same structure as the one in the DGD-RS update \eqref{Eq.DGD-RS-agentwise}, with the effective left-eigenvector scaling given by $\alpha_n/\gamma_n$.

In order to analyze the noisy consensus recursion, we define the estimation error of the decoded statistic $c_n[t]$ as
\begin{equation}
\epsilon_n[t] = \gamma_n \cdot \left(c_n[t] - \mathbb{E}\left[c_n[t] \,\vert\,  \mathcal{F}_t \right]\right), 
\end{equation} 
which satisfies $\mathbb{E}\left[\epsilon_n[t] \,\vert\,  \mathcal{F}_t \right] = 0$.
By adding and then subtracting $\zeta[t] \gamma_n \mathbb{E}\left[c_n[t] \,\vert\,  \mathcal{F}_t \right]$ in \eqref{Eq.UpdatedRule1}, the iterate can be re-written as
\begin{flalign}	
\label{Eq.UpdatedRule2}	
&x_{n}[t \! + \! 1] \\  
&=\! \mathcal{P}_{\Omega_n}\!\big[x_n[t] \!+\! \zeta[t] \gamma_n \mathbb{E}\! \left[c_n[t] \vert  \mathcal{F}_t \right] \!+\! \zeta[t] \epsilon_n[t] \!-\! \frac{\eta[t] h_{\bm\alpha,\bm\gamma}\gamma_n}{\alpha_n}d_n[t]\big],  \nonumber \\
&=\! \mathcal{P}_{\Omega_n} \! \Big[  \big( 1 \!-\! \zeta[t] \gamma_n \! \sum\limits_{\substack{m=1 \\ m\neq n}}^{N} \! \beta_{n,m} \alpha_{m}\big) x_n[t] \!+\! \zeta[t] \epsilon_n[t] \notag \\
& \  + \zeta[t] \gamma_n \! \sum\limits_{\substack{m=1 \\ m\neq n}}^{N} \! \beta_{n,m} \alpha_{m} x_m[t] - \underbrace{ \frac{\eta[t]h_{\bm\alpha,\bm\gamma}\, \gamma_n}{\alpha_n} \Big(x_{n}[t] \!-\! x_{n}[0]\Big)}_{\text{local-gradient correction}} \Big]. &&\nonumber
\end{flalign}
For explicit analysis, we rewrite \eqref{Eq.UpdatedRule2} into a matrix form. To this end, we first introduce 
\begin{equation}
\label{Eq.q}	
q_n=\frac{\alpha_n}{\gamma_n},\ 
\mathbf{q}=[q_1,\ldots,q_N]^\top, \ 
h_{\mathbf q} = \frac{N} {\sum_{n=1}^{N} \left(1/q_n\right)} = h_{\bm\alpha,\bm\gamma}. \nonumber
\end{equation}
Then, we define a weighted Laplacian matrix $\widetilde{\mathbf{L}}$ with the $(n,m)$-th entry given as
\begin{align}
\label{Eq.LaplacianMatrix}
l_{n, m} = \begin{cases}
- \gamma_n \cdot \beta_{n,m} \cdot \alpha_{m}, &n \neq m.  \\
\gamma_n \sum\limits_{\substack{k=1 \\ k\neq n}}^{N}\beta_{n,k} \cdot \alpha_{k}, & \text{else}.
\end{cases}
\end{align}

\begin{remark}
\label{remark2}		
The elements in the weighted Laplacian are determined by the transmit-power control coefficients, large-scale fading, and receive-scaling factors. 
By tuning these coefficients, we can modify, and potentially optimize the physical-layer connectivity between the agents to achieve faster in-network information combining. 
\end{remark}
We further define an OTA-induced mixing matrix $\mathbf{W}[t] = \mathbf{I}_N - \zeta[t] \widetilde{\mathbf{L}}$. 
Note that $\mathbf{W}[t]$ is deterministic and varies with $t$ only through the consensus stepsize $\zeta[t]$. 
In Section \ref{sec: Algorithm Analysis: Spectral Properties and Convergence Behavior}, we will show that, under the large-scale fading reciprocity (i.e., $\beta_{m,n}=\beta_{n,m}$), $\mathbf q$ is a positive left eigenvector of $\mathbf{W}[t]$ associated with eigenvalue $1$.
We collect the scaled/corrected local gradients in a vector 
\begin{equation}
    \mathbf d_{\mathbf q}[t]=\left[\frac{d_1[t]}{q_1},\ldots, \frac{d_N[t]}{q_N}\right]^\top,
\end{equation}
and the agent-wise estimation errors in a vector $\boldsymbol{\epsilon}[t]=[\epsilon_1[t],\ldots,\epsilon_N[t]]^\top$.
For the stacked state vector $\mathbf x[t]$, we define its joint feasible set as $\Omega \triangleq \Omega_1 \times \cdots \times \Omega_N = [-x_{\max},x_{\max}]^N$. 
Since each $\Omega_n$ is a compact interval, $\Omega$ is compact and convex.
Now \eqref{Eq.UpdatedRule2} can be written in matrix form as
\begin{equation}
\label{Eq.UpdatedRule3}
\mathbf{x}[t \! + \! 1]= \mathcal{P}_{\Omega}\Big[\underbrace{ \mathbf{W}[t] \cdot \mathbf{x}[t] - \eta[t] h_{\mathbf{q}} \cdot \mathbf{d_q}[t]}_{(a)} + \underbrace{\zeta[t]\cdot \bm\epsilon[t]}_{(b)} \Big].
\end{equation}

\begin{remark}
	\label{remark3}
	The term $(a)$ in \eqref{Eq.UpdatedRule3} follows the DGD-RS structure in \eqref{Eq.DGD-RS}. 
    It depends on the random $\mathbf{x}[t]$ but is $\mathcal F_t$-measurable. 
    Specifically, $\mathbf{W}[t]$ is row-stochastic, and $\mathbf d_{\mathbf q}[t]$ contains the local gradients scaled by the corresponding elements in the left eigenvector of $\mathbf{W}[t]$. 
    Compared with the DGD-RS algorithm in \eqref{Eq.DGD-RS}, our framework in \eqref{Eq.UpdatedRule3} further includes the projection operation and the aggregation noise captured by $\bm\epsilon[t]$.
    Hence, \eqref{Eq.UpdatedRule3} can be viewed as an implementation of decentralized projected gradient descent with row-stochastic mixing (DPGD-RS) via NCOTA aggregation over noisy fading channels.
\end{remark}

\begin{remark}
	\label{remark4}
	From a broader perspective, \eqref{Eq.UpdatedRule3} can be viewed as an average consensus iteration over noisy communication links. 
    The diminishing stepsize $\zeta[t]$ follows the design of stochastic approximation (SA) method to suppress the accumulation of communication disturbances, while the local-gradient correction $\eta[t] h_{\mathbf{q}} \mathbf{d_q}[t]$ acts as an anchoring term towards the agents' initial states. 
    As shown in our recent work \cite{deng2026exactaverageconsensusnoisy}, combining SA with such an anchoring mechanism enables exact average consensus under persistent disturbances. 
    This provides the underlying intuition of designing \eqref{Eq.UpdatedRule3}, whose convergence is established in Section \ref{sec: Algorithm Analysis: Spectral Properties and Convergence Behavior}.
\end{remark}

We refer to the iteration \eqref{Eq.UpdatedRule3} as the \textit{DPGD-RS-based average consensus (DPGD-RS-AC)} algorithm. The complete procedure is summarized in Algorithm \ref{alg:DPGD_RS_AC}: source-channel mapping $\rightarrow$ simultaneous NCOTA transmission $\rightarrow$ energy-based decoding $\rightarrow$ information combining $\rightarrow$ projected state updating with local-gradient correction. In this algorithm, the design variables are
\begin{itemize}
\item $\bm{\alpha}$: the transmit power-control vector;
\item $\bm{\gamma}$: the receive-scaling vector.
\end{itemize}
By tuning $\bm\alpha$ and $\bm\gamma$, the spectral properties of $\mathbf W[t]$ can be shaped to accelerate convergence, as shown in Section \ref{Power Control and Signal Scaling}.\footnote{Although the diminishing stepsize sequences also affect convergence, we prescribe them and focus on optimizing $\bm\alpha$ and $\bm\gamma$. Indeed, optimizing $\bm\gamma$ implicitly adjusts the effective consensus stepsize, since replacing $(\bm\gamma,\zeta[t])$ with $(c\bm\gamma,\zeta[t]/c)$ for any $c>0$ leaves the iteration unchanged.}

\begin{remark}
    The average channel gains $\{\beta_{m,n}\}$ need not be known for the implementation of NCOTA consensus, but their knowledge allows the transmit powers and receive-scaling coefficients to be optimized for faster convergence. 
    The only required channel-related parameter is the receiver noise power $\sigma^2$. 
    As noted in Remark \ref{remark: estimate error for sigma}, the convergence guarantee can be preserved even with an imperfect estimate of $\sigma^2$.
\end{remark}

\begin{algorithm}[h!]
\caption{DPGD-RS-AC via NCOTA Aggregation}
\small
\label{alg:DPGD_RS_AC}
\begin{algorithmic}[1]
\STATE \textbf{Initialize:} $\mathcal{Z}=\{-x_{\max},x_{\max}\}$, $\bm{\alpha}$, $\bm{\gamma}$, $\{\zeta[t]\}_{t\geq 0}$ and $\{\eta[t]\}_{t\geq 0}$.
\STATE Each agent $n\in\mathcal{N}$ initializes $x_n[0]\in\Omega_n$, computes $q_n$ and $h_{\mathbf q}$. 
\FOR{$t=0,1,\ldots, T\!-\!1$}
\FOR{each agent $n\in\mathcal{N}$}
\STATE Encode the local state $x_n[t]$ as \eqref{Eq.coding}.
\STATE Generate the transmit symbol $\mathbf{s}_n[t]$ as \eqref{Eq.transmitted_symbol}
\ENDFOR
\STATE All agents transmit and receive $\{\mathbf{s}_n[t]\}_{n\in\mathcal{N}}$ simultaneously over the shared wireless resource.
\FOR{each agent $n\in\mathcal{N}$}
\STATE Compute the combined symbol $c_n[t]$ by energy-based decoding as \eqref{Eq.update_vector1}.
\STATE Compute the local gradient $d_n[t]=\nabla f_n(x_n[t])=x_n[t]-x_n[0]$.
\STATE Update the local state $x_n[t\!+\!1]$ as \eqref{Eq.UpdatedRule1}.
\ENDFOR
\ENDFOR
\RETURN $\mathbf{x}[T]$.
\end{algorithmic}
\end{algorithm}

\subsection{Comparison to Existing Work}
A closely related work on NCOTA distributed consensus is \cite{10607888Yang}, where the proposed algorithm can be written as
\begin{equation}
\label{Eq.benchmark}
\mathbf{x}[t \! + \! 1]= \mathcal{P}_{\Omega}\left[ \mathbf{W}[t] \, \mathbf{x}[t] + \zeta[t] \, \bm\epsilon[t] \right].
\end{equation}
We refer to it as the \textit{plain average consensus} (PAC) protocol.
Compared with the proposed DPGD-RS-AC algorithm in \eqref{Eq.UpdatedRule3}, PAC does not include the local gradient term. Due to the channel noise and projection operation, the average is no longer preserved in every iteration. As a result, \eqref{Eq.benchmark} may reach consensus at a randomly biased value rather than $x^*$. This random bias issue of consensus averaging with noisy communication links has been thoroughly investigated in \cite{5411807LiTao}.

In contrast, \eqref{Eq.UpdatedRule3} forms a noisy gradient-based update that does not rely on preserving the network average at every iteration. 
In our recent work \cite{deng2026exactaverageconsensusnoisy}, we show that the local-gradient correction provides an anchoring mechanism towards the agents' initial states, thereby suppressing noise-induced random drifting and enabling exact average consensus.

\section{Algorithm Analysis: Spectral Properties and Convergence Behavior}
\label{sec: Algorithm Analysis: Spectral Properties and Convergence Behavior}
\subsection{Spectral Properties of the Mixing Matrix}
We start with the analysis of the weighted Laplacian matrix defined in~\eqref{Eq.LaplacianMatrix} and the OTA-induced mixing matrix $\mathbf{W}[t] = \mathbf{I}_N - \zeta[t] \widetilde{\mathbf{L}}$ in \eqref{Eq.UpdatedRule3}. Recall that large-scale fading reciprocity is assumed, i.e., $\beta_{n,m} = \beta_{m.n}$. The eigen-structures of $\widetilde{\mathbf{L}}$ and $\mathbf{W}[t]$ are characterized in the following lemmas. 
\begin{lemma}
\label{Lemma. spec_of_L}
The weighted Laplacian $\widetilde{\mathbf{L}}$ is generally non-symmetric unless $\gamma_n\alpha_m = \gamma_m\alpha_n$ for all $n,m \in \mathcal{N}$. 
Moreover, $\widetilde{\mathbf L}$ is similar to a symmetric positive semidefinite matrix 
\begin{equation}
\widetilde{\mathbf{L}}^{\operatorname{sym}} = \mathbf{Q}^{1/2} \widetilde{\mathbf{L}} \mathbf{Q}^{-1/2},
\end{equation}
where $\mathbf{Q} =  \operatorname{diag}(\mathbf{q})$ and $\mathbf{q} \triangleq [\alpha_{1}/\gamma_{1},\, \ldots,\, \alpha_{N}/\gamma_{N}]^{\top}$. $\widetilde{\mathbf{L}}$ and  $\widetilde{\mathbf{L}}^{\operatorname{sym}}$ share the same $N$ real eigenvalues that satisfy 
\begin{equation}
0 = \rho_1 < \rho_2 \leq \ldots \leq \rho_N \leq 2d_{\max},
\end{equation}
where $d_{\max} \triangleq \max\limits_{n\in\mathcal N} \sum\nolimits_{m \in \mathcal{N},\, m\neq n}\gamma_n \beta_{n,m}\alpha_m $.

\begin{proof}
Under the large-scale fading reciprocity, the symmetry condition of $\widetilde{\mathbf L}$ follows directly from its off-diagonal entries. 
In addition, $\mathbf Q \widetilde{\mathbf L}$ is symmetric as its $(n,m)$-th entry is $[\mathbf Q \widetilde{\mathbf L}]_{n,m} = -\alpha_n\beta_{n,m}\alpha_m$ for all $n,m\in\mathcal{N}$ and $n \neq m$. 
Hence, $\widetilde{\mathbf{L}}^{\operatorname{sym}} = \mathbf{Q}^{-1/2} (\mathbf Q \widetilde{\mathbf L})\mathbf{Q}^{-1/2} = \mathbf{Q}^{1/2} \widetilde{\mathbf{L}} \mathbf{Q}^{-1/2}$ is symmetric and similar to $\widetilde{\mathbf{L}}$, so they share the same $N$ real eigenvalues.

Applying the Gershgorin disc theorem to the rows of $\widetilde{\mathbf{L}}$, whose diagonal entries and absolute off-diagonal row sums are both given by $d_n = \sum\nolimits_{m \in \mathcal{N},\, m\neq n} \gamma_n \beta_{n,m}\alpha_m$, and noting that all eigenvalues are real, we obtain $0 \leq \rho_i \leq 2d_{\max}$ for all $i \in [N]$.
Therefore, $\widetilde{\mathbf{L}}^{\operatorname{sym}}$ is positive semidefinite.
Since $\widetilde{\mathbf{L}}^{\operatorname{sym}} \, \mathbf{Q}^{1/2} \, \mathbf{1} = \mathbf{0}$, the smallest eigenvalue is $\rho_1 = 0$. Since the underlying graph is complete with positive link weights, $\rho_1 = 0$ is simple and $\rho_2 > 0$.
\end{proof}
\end{lemma}

\begin{lemma}
\label{Lemma. spec_of_W}
The OTA-induced time-varying mixing matrix $\mathbf W[t]$ has unit row sums: $\mathbf W[t]\mathbf 1_N=\mathbf 1_N$. 
If $\zeta[t]\, d_{\max}\le 1$, then $\mathbf W[t]$ is nonnegative and hence row-stochastic. Under the large-scale fading reciprocity $\beta_{n,m}=\beta_{m,n}$, it holds that
\begin{equation}
\mathbf{q}^{\top} \widetilde{\mathbf{L}} = \mathbf{0}^{\top} \  		\Longrightarrow \  \mathbf{q}^{\top} \mathbf{W}[t]	= \mathbf q^{\top}(\mathbf I_N - \zeta[t]\widetilde{\mathbf L}) =\mathbf{q}^{\top} , \nonumber
\end{equation}
so $\mathbf q$ is a strictly positive left eigenvector of $\mathbf{W}[t]$ associated with eigenvalue $1$. 
\end{lemma}

The eigenvalues of $\mathbf W[t]$ are one-to-one determined by those of $\widetilde{\mathbf L}$. Specifically, $\lambda_i(\mathbf W[t]) = 1 - \zeta[t] \rho_i$ for all $i \in [N]$. 
Based on Lemma~\ref{Lemma. spec_of_L}, we can show that
\begin{equation}
\label{Eq.spec_bounds}
1 - 2\zeta[t]d_{\max} \leq \lambda_i(\mathbf W[t])  \leq 1.
\end{equation}
If $\bm{\alpha}$ and $\bm{\gamma}$ are selected to ensure $d_{\max} < 1/2$ and $\zeta[t] \leq 1$ for all $t \in \mathbb{N}_0$, then
\begin{equation}
\label{Eq.spec_bounds2}
0 \! = \! \rho_1 \!<\! \rho_i \!<\! 1,
\ 
\bigl|\lambda_i(\mathbf W[t])\bigr| \!<\! \bigl|\lambda_1(\mathbf W[t])\bigr| \!=\! 1, 
\  
\forall n \!\in\! [N] \backslash \{1\}. \nonumber
\end{equation}		

Although $\mathbf W[t]$ varies over time because of the stepsize $\zeta[t]$, its left eigenvector $\mathbf q$ is independent of $t$. Hence, once $\bm{\alpha}$ and $\bm{\gamma}$ are fixed, the extra coefficient $\{1/q_n\}_{n\in\mathcal N}$ used in the local-gradient correction are also fixed. 
By Lemma~\ref{Lemma. spec_of_W}, $\mathbf W[t]$ satisfies conditions \textup{(S1)--(S2)} in Lemma~\ref{Lemma.1} with $\mathbf u=\mathbf q$, while \textup{(S3)} additionally follows from the aforementioned eigenvalue bounds under $d_{\max}<1/2$ and $0<\zeta[t]\le1$.
Therefore, $\mathbf W[t]$ contracts the disagreement component in $\operatorname{ker}\{\mathbf q^\top\}$ while preserving the $\mathbf q$-weighted aggregate, thereby driving the iteration to a $\mathbf q$-weighted consensus value. The gradient correction step (scaled by left eigenvector) compensates for this imbalance by ensuring that $\mathbf q^\top\mathbf {d_q}[t]=\sum_{n=1}^{N}d_n[t]$, and thus recovers the aggregate gradient of the average-consensus objective.

\begin{remark}
When $q_m = q_n$ holds for all $m,\, n \in\mathcal N$, the induced $\mathbf{W}[t]$ becomes symmetric and doubly stochastic, and the term $(a)$ in \eqref{Eq.UpdatedRule3} reduces to the DGD structure considered in \cite{michelusi2024ota}. 
Our design relaxes this restriction, thereby enlarging the design space for optimizing the power control and scaling factors ($\boldsymbol{\alpha}$ and $\boldsymbol{\gamma}$) to achieve faster convergence.
\end{remark}

\subsection{Convergence Analysis}
Due to the randomness caused by noise and non-coherent interference, we consider the convergence of our proposed design in the mean-square sense. Particularly, we will show that the proposed DPGD-RS-AC algorithm \eqref{Eq.UpdatedRule3} achieves mean-square average consensus (MSAC), i.e.,
\begin{equation}
\label{Eq.MSAC}
\lim_{t\to\infty} \mathbb E\left[ \|\mathbf x[t]-\mathbf x^*\|^2 \right] =0,
\end{equation}
or equivalently,
$\lim_{t\to\infty}\mathbb E[(x_n[t]-x^*)^2]=0$ for all $n\in\mathcal N$. 
The convergence analysis follows a similar line of arguments to \cite{michelusi2024ota}, but extends it in one essential aspect: instead of relying on a symmetric doubly stochastic mixing matrix, we allow an OTA-induced \textit{row-stochastic} mixing matrix and compensate for the resulting imbalance through a \textit{left-eigenvector-scaled local-gradient correction} $h_{\mathbf q}\mathbf d_{\mathbf q}[t]$. 

To account for this extension and study the convergence of the DPGD-RS-AC algorithm \eqref{Eq.UpdatedRule3}, we first focus on the term (a), i.e., the error-free update, and temporarily ignore the estimation error in term (b). Let $\mathbf Q=\operatorname{diag}(\mathbf q)$. Since $\mathbf d_{\mathbf q}[t]=\mathbf Q^{-1}\nabla f(\mathbf x[t])$, the term (a) in \eqref{Eq.UpdatedRule3} can be rearranged to reveal a scaled gradient step with respect to an underlying time-varying Lyapunov energy. This energy captures both the consensus disagreement and the optimization objective in \eqref{Eq.DGD_problem}: 
\begin{align}
\label{Eq.deterministic update}
& \mathbf{W}[t]\mathbf{x}[t] - \eta[t]\, h_{\mathbf{q}} \cdot \mathbf{d_q}[t] \nonumber \\
& = \mathbf{x}[t] - \zeta[t]\widetilde{\mathbf{L}}\mathbf{x}[t]
- \eta[t]\, h_{\mathbf{q}} \, \mathbf{Q}^{-1}\nabla f(\mathbf{x}[t]) \nonumber \\
& = \mathbf{x}[t] - \eta[t] \, \underbrace{h_{\mathbf{q}} \, \mathbf{Q}^{-1} \left(\frac{\zeta[t]}{\eta[t]\, h_{\mathbf{q}}} \mathbf{Q} \widetilde{\mathbf{L}}\mathbf{x}[t]
+ \nabla f(\mathbf{x}[t])\right)}_{\text{scaled gradient of system energy}},
\end{align}
where scaling factor $h_{\mathbf{q}} \mathbf{Q}^{-1}$ comes from the scaled local-gradient correction in the DPGD-RS-AC update \eqref{Eq.UpdatedRule3}.
Using Lemma \ref{Lemma. spec_of_L}, we have $\widetilde{\mathbf{L}}^{\operatorname{sym}} = \mathbf{Q}^{1/2} \widetilde{\mathbf{L}} \mathbf{Q}^{-1/2}$ and hence $\mathbf{Q}\widetilde{\mathbf{L}} = \mathbf{Q}^{1/2} \widetilde{\mathbf{L}}^{\operatorname{sym}} \mathbf{Q}^{1/2}$.
Here, $\mathbf{Q}\widetilde{\mathbf{L}}$ is symmetric positive semidefinite.
Therefore, $\nabla_{\mathbf{x}}(\frac{1}{2}
\mathbf{x}^{\top}\mathbf{Q}\widetilde{\mathbf{L}}\mathbf{x}) = \mathbf{Q}\widetilde{\mathbf{L}}\mathbf{x}$.
Let us construct the following time-varying Lyapunov function:
\begin{equation}
\label{Eq.G}
G_t(\mathbf x)
\triangleq
\frac{\zeta[t]}{2\eta[t]h_{\mathbf q}} \mathbf x^{\top} \mathbf Q\widetilde{\mathbf L} \mathbf x + f(\mathbf x),
\end{equation}
then its gradient is given by
\begin{equation}
\label{Eq.nabla_G}
\nabla G_t(\mathbf x) = \frac{\zeta[t]}{\eta[t]h_{\mathbf q}} \mathbf Q\widetilde{\mathbf L}\mathbf x + \nabla f(\mathbf x),
\end{equation}
which is exactly the scaled gradient of system energy in \eqref{Eq.deterministic update}. The Lyapunov function $G_t(\mathbf x)$ contains two components: The first term measures the disagreement energy of $\mathbf x$ in the complement of the consensus subspace $\operatorname{ker}\{\mathbf{Q}\widetilde{\mathbf{L}}\}=\operatorname{span}\{\mathbf{1}\}$, while the second term represents the optimization objective $f(\mathbf x)$ defined in \eqref{Eq.DGD_problem}.
Finally, the DPGD-RS-AC algorithm \eqref{Eq.UpdatedRule3} can be re-written as
\begin{equation}
\label{Eq.UpdatedRule4}
\mathbf{x}[t \! + \! 1]= \mathcal{P}_{\Omega}\Big[\underbrace{\mathbf{x}[t] - \eta[t]\, h_{\mathbf{q}} \mathbf{Q}^{-1} \nabla G_t(\mathbf{x}[t])}_{(a)} + \underbrace{\zeta[t] \bm\epsilon[t]}_{(b)} \Big].
\end{equation}

\textit{MSAC Analysis:} We now analyze the convergence of \eqref{Eq.UpdatedRule4}. Note that this analysis is carried out for iterations $t \geq \tau$, where $\tau$ denotes an index after which the stepsizes $\zeta[t]$ and $\eta[t]$ become sufficiently small. To quantify convergence, we consider the following root mean-square consensus error:
\begin{equation}
\label{Eq. root mean-square CE}
E_{\mathrm{ce}}[t]
\!\triangleq\!
\Big(\!\frac{1}{N} \mathbb{E}\big[\sum_{n \in \mathcal{N}} \left(x_n[t] - x^* \right)^2\big]\Big)^{1/2}
\!\! =\!
\frac{1}{\sqrt N}
\left\|
\mathbf x[t]\!-\! \mathbf x^*
\right\|_{\mathbb E},
\end{equation}
where $\|\mathbf a\|_{\mathbb E} \triangleq \sqrt{ \mathbb E [ \|\mathbf a\|^2]}$ for any random vector $\mathbf a$. Indeed, $E_{\mathrm{ce}}[t]$ measures the average deviation of the local state value $x_n[t]$ from the global optimum $x^*$.

Next, we proceed by deriving an explicit upper bound on $E_{\mathrm{ce}}[t]$.
Two auxiliary sequences will be introduced. First, we define $\tilde{\mathbf{x}}[t]$ as a noise-free iterate generated by removing the estimation error from \eqref{Eq.UpdatedRule4}, i.e., 
\begin{equation}
\label{Eq.Noise-Free-State}
\tilde{\mathbf{x}}[t \! + \! 1] = \mathcal{P}_{\Omega}\left[\tilde{\mathbf{x}}[t] - \eta[t] h_{\mathbf{q}} \mathbf{Q}^{-1} \nabla G_t(\tilde{\mathbf{x}}[t])\right], \  \forall  t\geq\tau,
\end{equation}
with $\tilde{\mathbf{x}}[t]=\mathbf{x}[t]$ for $t<\tau$. Second, we define $\mathbf{x}^{\dagger}[t]$ as the instantaneous minimizer of the time-varying Lyapunov function $G_t$ over the feasible set, i.e., 
\begin{equation}
\label{Eq.Local-optimum-State}
\mathbf{x}^{\dagger}[t] \triangleq	\underset{\mathbf{x}\in\Omega}{\arg\min} G_t(\mathbf{x}).
\end{equation}
Substituting $\tilde{\mathbf{x}}[t]$ and $\mathbf{x}^{\dagger}[t]$ into \eqref{Eq. root mean-square CE} yields a decomposition as
\begin{align}
\label{Eq.ErrorBound}	
& E_{\mathrm{ce}}[t] \nonumber \\
& \triangleq \frac{1}{\sqrt N} \left\| \mathbf x[t]-\mathbf x* \right\|_{\mathbb E}  \nonumber\\
& = \frac{1}{\sqrt N} \left\| \mathbf{x}[t] - \tilde{\mathbf{x}}[t] + \tilde{\mathbf{x}}[t] - \mathbf{x}^{\dagger}[t] + \mathbf{x}^{\dagger}[t]- \mathbf{x}^{*} \right\|_{\mathbb{E}}  \\
& \stackrel{(c)}{\leq} \underbrace{\frac{1}{\sqrt N} \left\| \mathbf{x}[t] - \tilde{\mathbf{x}}[t]  \right\|_{\mathbb{E}}}_{(d)} +  \underbrace{\frac{1}{\sqrt N} \left\| \tilde{\mathbf{x}}[t] - \mathbf{x}^{\dagger}[t]\right\|_{\mathbb{E}}}_{(e)} \nonumber \\
& \quad + \underbrace{ \frac{1}{\sqrt N}  \left\| \mathbf{x}^{\dagger}[t] - \mathbf{x}^{*} \right\|}_{(f)}, \nonumber
\end{align}
where the inequality $(c)$ follows from the Minkowski inequality \cite[Lemma $14.10$]{FlorescuTudor2013}.

The decomposition in \eqref{Eq.ErrorBound} separates the sources of error in the convergence analysis. Term $(d)$ measures the error between the actual stochastic iterate $\mathbf{x}[t]$ and the noise-free iterate $\tilde{\mathbf{x}}[t]$, and hence captures the accumulated effect of the non-coherent interference and aggregated noise. Term $(e)$ captures the error between $\tilde{\mathbf{x}}[t]$ and the instantaneous optimizer $\mathbf{x}^{\dagger}[t]$ of $G_t(\cdot)$. Here, $\mathbf{x}^{\dagger}[t]$ serves as a time-varying intermediate target: the Lyapunov function $G_t(\cdot)$ gradually strengthens the penalty on state disagreement as the ratio $\zeta[t]/\eta[t]$ increases, thereby driving $\mathbf{x}^{\dagger}[t]$ toward the exact average-consensus vector $\mathbf{x}^{*}$. Term $(f)$ measures the remaining distance from $\mathbf{x}^{\dagger}[t]$ to $\mathbf{x}^{*}$. 

To bound term $(d)$, we first establish a uniform bound on the conditional second moment of the estimation error $\bm{\epsilon}[t]$.
\begin{lemma}
\label{Lemma.4}
The second-order moment of $\bm\epsilon[t]$, conditioned on $\mathcal{F}_t$, is upper bounded:
\begin{align}
\label{Eq.error_budget}
&\frac{1}{N}\mathbb{E}\left[\| \bm\epsilon[t] \|^2  \vert \mathcal{F}_t \right] \nonumber \\ 
&\leq \Sigma = \frac{1}{N}\sum_{n=1}^{N}\Big\{2x_{\max} \gamma_n\Big(\sum\limits_{\substack{m=1 \\ m\neq n}}^{N} \beta_{n,m}\alpha_{m} + 2\sigma^2 \Big) \Big\}^2. 
\end{align}
Proof. See Appendix A.
\end{lemma}

Lemma \ref{Lemma.4} implies $\mathbb{E}[ \|\zeta[t]\bm\epsilon[t]\|^2\,\vert\,\mathcal{F}_t]\leq \zeta^2[t]N\Sigma$. Since $\zeta[t] \!\to\! 0$ and  $\sum_{t=0}^{\infty}\zeta^2[t]<\infty$, both the instantaneous and accumulated mean-square effects of $\bm{\epsilon}[t]$ remain controlled. This provides the key ingredient for bounding term $(d)$.

We are now ready to state the main convergence result. 
Theorem \ref{Theorem.1} provides explicit upper bounds for the three terms in \eqref{Eq.ErrorBound}, all of which vanish asymptotically, thereby establishing MSAC of the proposed DPGD-RS-AC algorithm \eqref{Eq.UpdatedRule4}.

\begin{theorem}
\label{Theorem.1}
Let $\zeta[t] = \zeta[0]\left(1+\frac{t}{\delta}\right)^{-3/4}$ and $\eta[t] = \eta[0]\left(1+\frac{t}{\delta}\right)^{-1}$, where $\delta \geq \frac{5}{4 \eta[0]}$. Assume the following constraints are satisfied for all $t \geq \tau$: $\mathbf{(\romannumeral 1)}$ $2h_{\mathbf{q}} \eta[t] + \zeta[t] \rho_N \leq 2$; $\mathbf{(\romannumeral 2)}$ $\frac{\eta[t]}{\zeta[t]} \leq \frac{\Delta \cdot \rho_2}{\sqrt{N} \cdot h_{\mathbf{q}} \cdot \| \mathbf{Q}^{-1} \| \cdot \nabla^*}$; $\mathbf{(\romannumeral 3)}$ $\frac{\eta[t]}{\zeta[t]} \geq \frac{\eta[t \! + \! 1]}{\zeta[t \! + \! 1]} $. Then, for the distributed optimization problem \eqref{Eq.DGD_problem}, the following upper bounds hold:
\begin{align}
&\frac{1}{\sqrt{N}}\Big\|\mathbf{x}[t]-\tilde{\mathbf{x}}[t] \Big\|_{\mathbb{E}} \leq \frac{\sqrt{5} e}{2}\Big[\frac{\zeta[0]}{\sqrt{\eta[0]}} \sqrt{\Sigma}\Big] \Big(1+\frac{t}{\delta}\Big)^{-\frac{1}{4}}.   \label{Theorem.1.1} \\
&\frac{1}{\sqrt{N}}\Big\|\tilde{\mathbf{x}}[t]-\mathbf{x}^{\dagger}[t]\Big\|_{\mathbb{E}} \leq  \operatorname{dm}(\Omega_n) \Big(1 \!+\! \frac{t \!-\! \tau}{\tau \!+\! \delta}\Big)^{-\frac{5}{4}} \!\!+\! \Big(1 \!+\! \frac{t}{\delta}\Big)^{-\frac{1}{4}}  \nonumber \\ 
& \ \times \! \frac{\| \mathbf{Q}^{-1}\|  \nabla^*  e \eta[0] h_{\mathbf{q}}}{4  \rho_2  \zeta[0]} \Big( \!1 \!+\! \frac{\| \mathbf{Q}^{-1} \|  \eta[0] h_{\mathbf{q}} }{\rho_2  \zeta[0]}\Big(\! 1\!+\!\frac{t}{\delta}\Big)^{-\frac{1}{4}}\Big). \label{Theorem.1.2} \\
& \frac{1}{\sqrt{N}}\Big\|\mathbf{x}^\dagger[t] - \mathbf{x}^*\Big\| \leq \frac{\| \mathbf{Q}^{-1}\|  \nabla^*  \eta[0] h_{\mathbf{q}}}{\rho_2 \zeta[0]} \left(1+\frac{t}{\delta}\right)^{-\frac{1}{4}}. \label{Theorem.1.3}
\end{align}
\end{theorem}

Here, $\Delta \triangleq \min_{x\in\operatorname{bd}(\Omega_n)}|x-x^*|$ is the distance from $x^*$ to the boundary of $\Omega_n$, and $\nabla^*\triangleq\max_{n\in \mathcal N}|\nabla f_n(x^*)|$ is the maximum magnitude of the local gradients evaluated at $x^*$.
\begin{proof}
The proof follows a similar line of argument to that of \cite[Theorem 4]{michelusi2024ota}, with adaptations to allow for a heterogeneous stepsize applied to the gradient-descent terms instead of a global stepsize $\eta[t]$ shared by all agents. Specifically, in the deterministic part of \eqref{Eq.UpdatedRule4}, the weighted gradient term $\eta[t]\, h_{\mathbf{q}} \, \mathbf{Q}^{-1} \cdot \nabla G_t(\mathbf{x}[t])$ can be equivalently re-written by using a diagonal heterogeneous stepsize matrix 
\begin{equation}
\bm\Lambda[t] \triangleq \eta[t] \, h_{\mathbf{q}} \, \mathbf{Q}^{-1} = \operatorname{diag}\left(\frac{\eta[t]\, h_{\mathbf{q}}}{q_1},\ldots,\frac{\eta[t]\, h_{\mathbf{q}}}{q_N}\right).
\end{equation}
Hence, the update can be viewed as a projected gradient descent with heterogeneous stepsizes.
\begin{equation}
\label{Eq.UpdatedRule5}
\mathbf{x}[t \! + \! 1]= \mathcal{P}_{\Omega}\Big[\mathbf{x}[t] - \bm\Lambda[t] \nabla G_t(\mathbf{x}[t]) + \zeta[t] \bm\epsilon[t] \Big].
\end{equation}
Since $\bm\Lambda[t]$ is diagonal and positive definite, its induced norm satisfies
\begin{equation}
\| \bm\Lambda[t] \| = \max_{n} \eta[t] \, h_{\mathbf{q}}/q_n = \eta[t] \, h_{\mathbf{q}} \, \|\mathbf{Q}^{-1}\| \geq \eta[t].
\end{equation} 

Therefore, the heterogeneous local-gradient correction can be conservatively bounded by replacing the scalar stepsize in \cite[Theorem 4]{michelusi2024ota} with the maximum effective stepsize $\eta[t]h_{\mathbf q}\|\mathbf Q^{-1}\|$. 
Under conditions $\mathbf{(\romannumeral 1)}$--$\mathbf{(\romannumeral 3)}$, the same contraction and perturbation arguments then yield the three bounds in \eqref{Theorem.1.1}--\eqref{Theorem.1.3}.
\end{proof}

By Theorem \ref{Theorem.1}, the expected squared distance between $\mathbf{x}[t]$ and $\mathbf{x}^{*}$ is upper bounded as:
\begin{align}
\label{Eq.bound}
& \mathbb{E}\left[\| \mathbf{x}[t] \!-\! \mathbf{x}^{*} \|^2\right] \\
& \leq\! N \Big(\frac{5 e^2\zeta[0]^2\Sigma}{4\eta[0]}\ \!\!+\!\! \Big(\!\frac{(4\!+\! e) \| \mathbf{Q}^{-1}\| \nabla^*  \eta[0] h_{\mathbf{q}}}{4\rho_2 \zeta[0]}\Big)^2 \Big) \Big(1\!+\!\frac{t}{\delta}\Big)^{-\frac{1}{2}}, \nonumber 
\end{align}
implying $\lim_{t\to\infty} \mathbb{E} \big[ \left\| \mathbf{x}[t]-\mathbf{x}^{\star} \right\|^2 \big]=0$. Thus, the DPGD-RS-AC update given in \eqref{Eq.UpdatedRule4} converges in mean square sense to the exact average consensus vector $\mathbf{x}^{*}$.

\section{Power Control and Signal Scaling}
\label{Power Control and Signal Scaling}

In the previous section, we show that the convergence speed of our proposed DPGD-RS-AC algorithm \eqref{Eq.UpdatedRule4} depends on three key quantities: (\romannumeral 1) the estimation-error budget $\Sigma$, which upper-bounds the conditional mean-square estimation error caused by non-coherent interference and  noise; (\romannumeral 2)  $\rho_2$: the second smallest eigenvalue of the weighted Laplacian $\widetilde{\mathbf L}$; and (\romannumeral 3) the harmonic-mean normalized factor $h_{\mathbf q} \|\mathbf Q^{-1}\|$, which quantifies the imbalance introduced by the local-gradient correction. Since these quantities all depend on the transmit power control vector $\bm\alpha$ and the receive scaling vector $\bm\gamma$, they are inherently coupled. Improving the convergence speed of the DPGD-RS-AC algorithm \eqref{Eq.UpdatedRule4} requires a joint design of $\bm\alpha$ and $\bm\gamma$ that optimizes these coupled quantities.

\subsection{Problem formulation}
We aim at optimizing $\bm\alpha$ and $\bm\gamma$ by minimizing the leading coefficient of \eqref{Eq.bound}. The resulting optimization problem is formulated as
\begin{align}
\tag{P0}\label{P0}	
\min_{\bm{\alpha,\, \bm\gamma}} \quad & 
C_1 \cdot \Sigma(\bm\alpha, \bm\gamma) + C_2^2 \cdot \frac{\| \mathbf{Q}^{-1}(\bm\alpha, \bm\gamma)\|^2}{\rho_2(\bm\alpha, \bm\gamma)^2 \cdot (\operatorname{tr}({\mathbf{Q}^{-1}}))^{2}}, \\
\text { s.t. } 			\quad & 0<\alpha_n \leq \alpha_{\max }, \  0<\gamma_n \leq \gamma_{\max },  \  \forall n\in \mathcal{N}, \nonumber \\
& \rho_N(\bm\alpha, \bm\gamma) < 1. \nonumber
\end{align}
where $C_1 \!=\! 5e^2 \zeta[0]^2 \!/(4 \eta[0])$ and $C_2 \!=\! (4\!+\!e)  N\nabla^* \eta[0] /\! (4\zeta[0])$, and $h_{\mathbf q} \triangleq N / \sum_{n \in \mathcal{N}} \left(1/q_n\right) = \frac{N}{\operatorname{tr}({\mathbf{Q}^{-1}})}$. Note that $\| \mathbf{Q}^{-1} \| = \| \operatorname{diag}(\mathbf{q})^{-1}\| = \max_n{\frac{1}{q_n}} = \max_n{\frac{\gamma_n}{\alpha_n}}$ is non-smooth. 
Thus, we introduce an auxiliary variable $s$ through an epigraph reformulation, and replace the non-smooth $\| \mathbf{Q}^{-1} \|$ with $s$ subject to a smooth bilinear constraints $s \geq \frac{\gamma_n}{\alpha_n}$ for all $n\in\mathcal N$. 
This reformulation preserves optimality while facilitating numerical optimization.

Although \eqref{P0} follows directly from \eqref{Eq.bound}, its weighted-sum objective may be numerically sensitive because the two terms can differ substantially in magnitude. To better isolate the effects of the mixing matrix, we instead adopt an $\varepsilon$-constraint formulation, in which the estimation-error budget is imposed as a hard constraint while the objective focuses solely on the spectral convergence term. This leads to
\begin{align}
\tag{P1}\label{P1}	
\min_{\bm{\alpha},\, \bm\gamma,\, s} \quad &  \frac{s^2}{\rho_2(\bm\alpha, \bm\gamma)^2\cdot (\operatorname{tr}({\mathbf{Q}^{-1}}))^{2}}, \\
\text { s.t. } 			\quad & 0<\alpha_n \leq \alpha_{\max }, \quad 0<\gamma_n \leq \gamma_{\max }, \quad \forall n\in \mathcal{N}, \nonumber \\
& \gamma_n / \alpha_n \leq s,  \quad \forall n\in \mathcal{N},  \nonumber \\
& \rho_N(\bm\alpha, \bm\gamma) < 1, \quad \Sigma(\bm\alpha, \bm\gamma) \leq \bar\Sigma \nonumber,
\end{align}
where $\bar\Sigma>0$ is a prescribed estimation-error budget. \eqref{P1} can be interpreted as an $\varepsilon$-constraint reformulation of the same bi-objective trade-off originated from \eqref{P0}. Specifically, the estimation-error term $\Sigma(\bm\alpha,\bm\gamma)$ is constrained by $\bar\Sigma$, and the objective focuses on minimizing the spectral convergence term. In this way, \eqref{P1} separates the two competing components: the stochastic error caused by interference and noise is kept below a prescribed level, whereas the spectral term is optimized to accelerate consensus.

Note that both $\rho_2(\bm\alpha,\bm\gamma)$ and $\operatorname{tr}(\mathbf Q^{-1})$ in \eqref{P1} are non-linear and jointly non-convex functions of $\bm\alpha$ and $\bm\gamma$. In addition, the epigraph variable $s$ is coupled with them through the constraints $\gamma_n/\alpha_n\le s$. Moreover, the spectral constraint is determined by the largest eigenvalue $\rho_N$ of the OTA-induced Laplacian matrix $\widetilde{\mathbf{L}}$. Therefore, \eqref{P1} is a non-convex non-linear program (NLP) with spectral constraints. To improve the numerical tractability of \eqref{P1}, we replace the strict inequalities in \eqref{P1} with non-strict constraints by introducing small positive margins: $\alpha_{\min}$, $\gamma_{\min}$, and $\delta_{\rho}$. These margins enforce $\alpha_n \geq \alpha_{\min}$, $\gamma_n \geq \gamma_{\min}$, and $\rho_N(\bm\alpha,\bm\gamma)\leq 1-\delta_{\rho}$, respectively. Together with the upper bounds on $\bm\alpha$, $\bm\gamma$, and $s$, these constraints make the feasible set compact. The resulting reformulated problem is given by
\begin{align}
\tag{P2}\label{P2}	
& \min_{\bm{\alpha},\, \bm\gamma,\, s} \quad  g(\bm{\alpha},\, \bm\gamma,\, s) = \frac{s^2}{\rho_2(\bm\alpha, \bm\gamma)^2\cdot (\operatorname{tr}({\mathbf{Q}^{-1}}))^{2}}, \\
\text { s.t. }  
& \left(\mathbf{C1}\right)	\quad  \alpha_{\min} - \alpha_n \leq 0,  \quad  \alpha_n - \alpha_{\max} \leq 0, \quad \forall n \in \mathcal{N},  \nonumber \\
& \left(\mathbf{C2}\right)	\quad  \gamma_{\min} - \gamma_n \leq 0,  \quad  \gamma_n - \gamma_{\max} \leq 0, \quad \forall n \in \mathcal{N},  \nonumber \\
&\left(\mathbf{C3}\right)	\quad \frac{\gamma_n}{\alpha_n} - s \leq 0, \quad  s- \frac{\gamma_{\max}}{\alpha_{\min}}\leq 0,\quad \forall n \in \mathcal{N},  \nonumber \\
&\left(\mathbf{C4}\right)	\quad  \rho_N(\bm\alpha, \bm\gamma) - \left(1-\delta_{\rho}\right) \leq 0, \nonumber \\
&\left(\mathbf{C5}\right)	\quad 
\Sigma(\bm\alpha, \bm\gamma) - \bar{\Sigma} \leq 0. \nonumber
\end{align}

\subsection{Proposed Solution}
In this section, we propose a two-layer scheme for solving \eqref{P2}. The inner layer applies a constrained trust-region method \cite{nocedal2006numerical} to solve \eqref{P2} under a fixed estimation-error budget $\bar{\Sigma}$, while the outer layer monotonically tightens $\bar{\Sigma}$ to explore the trade-off. 
Specifically, at the $k$-th outer iteration, we update the estimation-error budget by $\bar{\Sigma}[k]=0.9\bar{\Sigma}[k-1]$. 
For each outer iteration, the solution is accepted only if it is feasible and yields a non-increasing objective value $g(\bm\alpha, \bm\gamma, s)$. Otherwise, the outer loop terminates and returns the last accepted solution. We refer to this power-control and receive-scaling algorithm as the spectral-reshaping power control (SR-PC) algorithm.

\begin{algorithm}[h!]
\caption{Spectral-Reshaping Power Control (SR-PC)}
\small
\label{alg:SRPC}
\begin{algorithmic}[1]
\STATE \textbf{Initialize:} budget $\bar{\Sigma}[0]$; optimal objective $g^*= +\infty$; counter $k=0$; feasible initial point $(\bm\alpha[0], \bm\gamma[0], s[0])$.
\STATE Define \eqref{P2} with $\bar{\Sigma}[0]$ and initialized by $(\bm\alpha[0], \bm\gamma[0], s[0])$. 
\STATE Solve \eqref{P2} using trust-region method: obtain $g(\bm\alpha^*, \bm\gamma^*, s^*)$ and minimizer $(\bm\alpha^*, \bm\gamma^*, s^*)$.
\WHILE {$(\bm\alpha^*, \bm\gamma^*, s^*)$ is feasible and $g^* \geq g(\bm\alpha^*, \bm\gamma^*, s^*)$} 
\STATE $k \gets k+1 $ 
\STATE $(\bm\alpha[k], \bm\gamma[k], s[k]) \gets (\bm\alpha^*, \bm\gamma^*, s^*)$
\STATE $g^* \gets g(\bm\alpha^*, \bm\gamma^*, s^*)$
\STATE $\bar{\Sigma}[k] \gets 0.9\bar{\Sigma}[k-1]$
\STATE Define problem \eqref{P2} with budget $\bar{\Sigma}[k]$ and initialized by $(\bm\alpha[k], \bm\gamma[k], s[k])$.
\STATE Solve \eqref{P2} using trust-region method: obtain $g(\bm\alpha^*, \bm\gamma^*, s^*)$ and minimizer $(\bm\alpha^*, \bm\gamma^*, s^*)$.
\ENDWHILE
\STATE $(\bm\alpha^*, \bm\gamma^*, s^*) \gets (\bm\alpha[k], \bm\gamma[k], s[k])$
\RETURN $\bm\alpha^*$, $\bm\gamma^*$, and $s^*$.
\end{algorithmic}
\end{algorithm}

The overall structure of the proposed SR-PC algorithm is summarized in Algorithm \ref{alg:SRPC}. 
A key property of \eqref{P2} is that the objective function and all constraint functions in $\left(\mathbf{C1}\right)$--$\left(\mathbf{C5}\right)$ admit explicit first-order derivatives in closed-form whenever the relevant eigenvalues $\rho_2$ and $\rho_N$ are simple. 
Here, we assume that they remain simple throughout the feasible region of \eqref{P2}.\footnote{For real-analytic symmetric matrix families without structurally enforced multiplicities, repeated eigenvalues occur only on a Lebesgue measure-zero parameter set and can typically be avoided by arbitrarily small perturbations.}
All the first-order information is supplied to the \textit{trust-constr} solver in SciPy for solving \eqref{P2} in the inner layer, while the required second-order information is approximated using the Broyden–Fletcher–Goldfarb–Shanno (BFGS) quasi-Newton method. 
The solver constructs a local quadratic model of the objective $g(\bm\alpha,\bm\gamma,s)$ within a trust region around the current iterate, together with linearized approximations of the constraints. Solving the resulting trust-region sub-problem provides a local improvement direction for updating the current iterate. 
Repeating this procedure generates an iterate sequence for updating the optimization variables $\bm\alpha$, $\bm\gamma$, and $s$. We assume that a standard linear independence constraint qualification (LICQ) assumption holds at all accumulation points of the iterate sequence. This is a generic nondegeneracy condition in constrained nonlinear programming.

With the eigenvalue simplicity, the objective and constraint functions of \eqref{P2} are real-analytic in a neighborhood of the compact feasible set. 
Consequently, they are globally Lipschitz continuous and Lipschitz smooth. 
These properties ensure uniform second-order approximation error bounds for the local linear–quadratic models employed by the constrained trust-region method.
Under LICQ and the standard convergence conditions of the constrained trust-region method \cite{nocedal2006numerical}, every accumulation point of the inner solver's iterate sequence satisfies the first-order Karush–Kuhn–Tucker (KKT) conditions.

\subsection{Alternative Benchmark Schemes}
\label{sec:alternative}

To evaluate the performance of our proposed SR-PC algorithm, we implement two benchmark schemes: \textit{full power (FP)} and \textit{spectral radius minimization (SRM)}. 

\subsubsection{Full power (FP) scheme}
In this scheme, we set $\bm\alpha = \alpha_{\max}\, \mathbf{1}$, so that all transmitters operate at their maximum power. In addition, the receiver gains are chosen uniformly as $\bm\gamma = \bar{\gamma}\, \mathbf{1}$, where $\bar{\gamma} =$ $1/\max_{n \in \mathcal{N}}\left\{\sum\limits_{m=1, m\neq n}^{N}\beta_{n,m} \cdot \alpha_{m}\right\}$. 
This choice ensures that every entry of $\widetilde{\mathbf{L}}$ is less or equal to $1$. By construction, the time-varying mixing matrix $\mathbf{W}[t]$ is doubly stochastic. Furthermore, when $\zeta[t]d_{\max}<1/2$, all eigenvalues of $\mathbf{W}[t]$ are positive, and $\rho(\mathbf{W}[t] - \frac{\mathbf{1} \mathbf{1}^{\top}}{N})=1-\zeta\rho_2<1$.
Therefore, $\mathbf{W}[t]$ satisfies the convergence conditions \textup{(S1)--(S3)} in Lemma \ref{Lemma.1}, which ensures that the FP scheme drives the NCOTA network to average consensus. Notably, FP is the most energy-intensive scheme that still satisfies the information-balance requirement: all agents transmit at full power and the receiver gains are uniformly scaled to yield balanced network interactions. 

\subsubsection{SRM scheme}

In a preliminary work \cite{icc25}, we adopted the following optimization design motivated by minimizing the spectral norm $\rho(\mathbf{W} - \mathbf{1} \mathbf{1}^{\top}/N)$:
\begin{align}
\label{Eq.SRM_SDP}	
\min_{\bm{\alpha},\, \bm{\gamma},\, S} \quad & \quad\quad S, \\
\text { s.t. } 			\quad & \begin{bmatrix} S \cdot I & \mathbf{W} - \mathbf{1} \mathbf{1}^{\top}/{N} \\ \mathbf{W}^{\top} - \mathbf{1} \mathbf{1}^{\top}/{N} & S \cdot I \end{bmatrix}   \succeq 0;  \nonumber \\
& 0 < \alpha_n \leq \alpha_{\max}, \ 0<\gamma_n \leq \gamma_{\max}, \quad \forall n\in \mathcal{N};  \nonumber \\
& \alpha_n\gamma_m = \alpha_m\gamma_n, \quad \forall n,m \in \mathcal{N}. \nonumber
\end{align}

The last constraint ensures that $\bm\alpha\propto\bm\gamma$, which makes $\mathbf W$ symmetric. 
Since $\mathbf W$ is row-stochastic by construction, it is also doubly stochastic.
If the optimal solution of \eqref{Eq.SRM_SDP} satisfies $S^*<1$, $\mathbf W$ satisfies the consensus conditions \textup{(S1)--(S3)} in Lemma~\ref{Lemma.1}, and hence the SRM scheme guarantees the convergence of the DPGD-RS-AC algorithm \eqref{Eq.UpdatedRule3} to average consensus. 
Indeed, the optimum necessarily satisfies $S^*<1$, since choosing $\bm\alpha=\bm\gamma = \upsilon \mathbf 1$ with sufficiently small $\upsilon>0$ yields a feasible solution satisfying this requirement.

The problem involves semidefinite constraints but is jointly non-convex in $\bm\alpha$ and $\bm\gamma$ due to the bilinear terms. 
Following our prior work \cite{icc25}, we solve \eqref{Eq.SRM_SDP} via alternating minimization (AM) that iteratively optimizes $\bm\alpha$ and $\bm\gamma$ while keeping the other fixed. As shown in \cite{icc25}, each subproblem becomes a convex semidefinite program (SDP) and admits a global minimizer that can be computed in polynomial time. 

\section{Simulation Results}
\label{Simulation Results}
We consider a network deployed over a rectangular area of $100\times100\,\mathrm{m^2}$, where $N = 18$ agents are uniformly and randomly deployed. To avoid unrealistic clustering, a minimum inter-agent distance of $d = 12\,\mathrm{m}$ is imposed. The large-scale fading coefficient $\beta_{n,m}$ of the link between agent $m$ and $n$ is determined by \cite[Eq.$\,$(2.52)]{goldsmith2005wireless}: 
\begin{align}
\label{Eq.large_scale_fading}	
\beta_{m, n}(\mathrm{dB}) = K(\mathrm{dB})-10 \xi \log _{10}\left[d_{m,n}/d_0\right] + \Psi(\mathrm{dB}),
\end{align} 
where $\xi = 3.5$ is the path-loss exponent, $d_{m,n}$ is the distance between agents $m$ and $n$, $d_0 = 10\,\mathrm{m}$ is the reference distance at which the path loss is $K = -65\,\mathrm{dB}$, and $\Psi$ models the shadowing effect, following $\mathcal{N}(0,\sigma_{\psi}^2)$ with $\sigma_{\psi} = 5\,\mathrm{dB}$. 
The receiver noise power is set to $-99.5\,\mathrm{dBm}$. 
For reference, a transmit power of $1\,\mathrm{mW}$ yields an SNR of approximately $10\,\mathrm{dB}$ at a distance of $50\,\mathrm{m}$, excluding shadowing.
The initial state values $\{x_{n}[0]\}_{n \in \mathcal{N}}$ are generated independently from a uniform distribution $\mathcal{U}\left[-500, \, 500\right]$ with $x_{\max} = 500$. 

In our simulations, we keep the gradient stepsize $\eta[t]$ as specified in Theorem \ref{Theorem.1} with $\eta[0] = 0.1$, while adopting a more slowly decaying consensus stepsize $\zeta[t]=\zeta[0](1+t/\delta)^{-0.6}$ with $\zeta[0] = 0.9$. 
Compared with the exponent $3/4$ in Theorem 1, the exponent $0.6$ retains stronger information mixing over the finite simulation horizon while satisfying the standard diminishing stepsize conditions. Moreover, the initial stepsizes are selected empirically to obtain stable convergence results.

\begin{figure}[h!]
\centering
\subfloat[DPGD-RS-AC (SR-PC)]{
\includegraphics[width=0.22\textwidth]{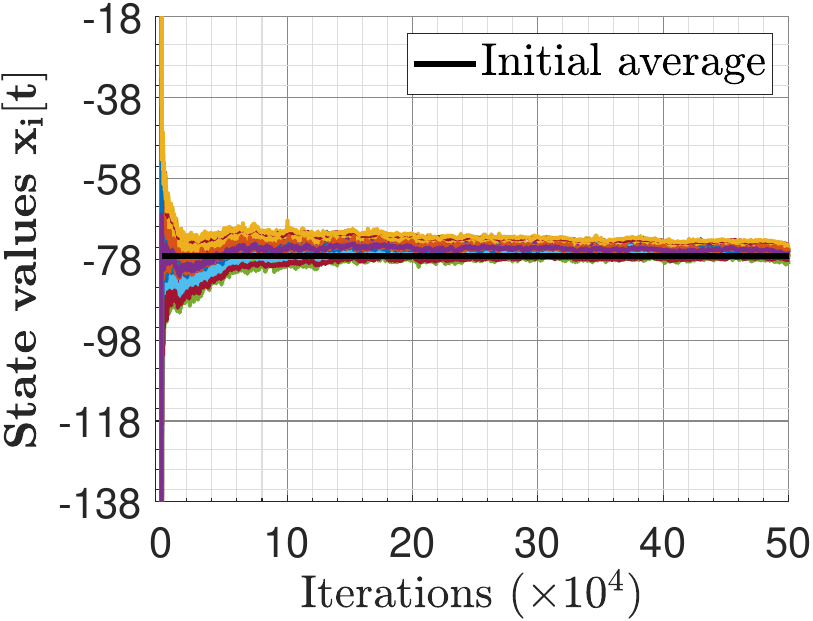}
\label{Fig.3:sub1}
}
\hspace{0pt}
\subfloat[PAC]{
\includegraphics[width=0.22\textwidth]{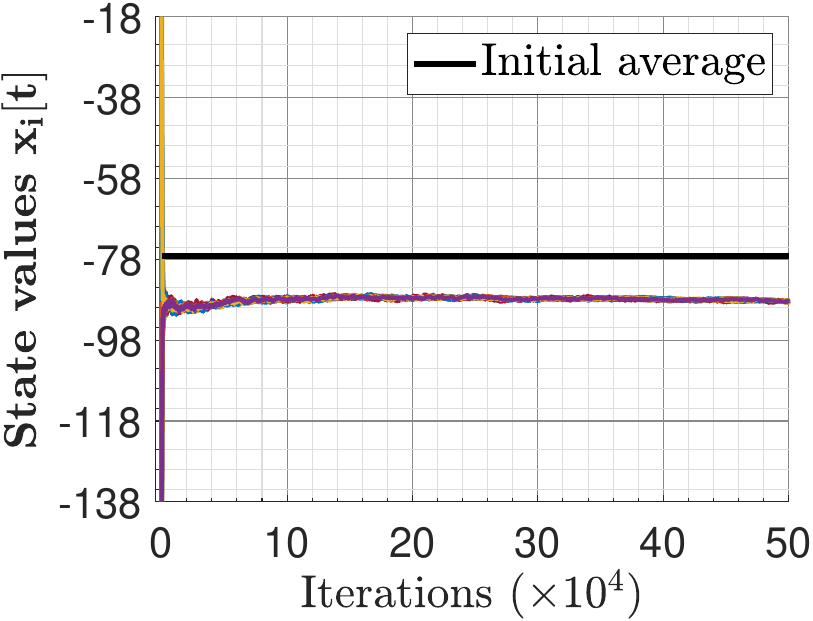}
\label{Fig.3:sub3}
}\\
\subfloat[DPGD-RS-AC (SRM)]{
\includegraphics[width=0.22\textwidth]{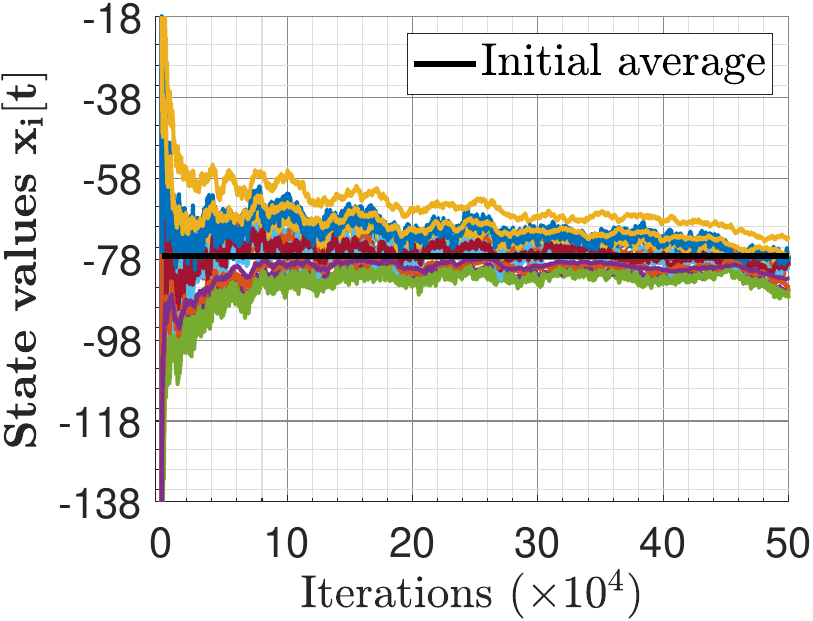}
\label{Fig.3:sub2}
}
\hspace{0pt}
\subfloat[DPGD-RS-AC (FP)]{
\includegraphics[width=0.22\textwidth]{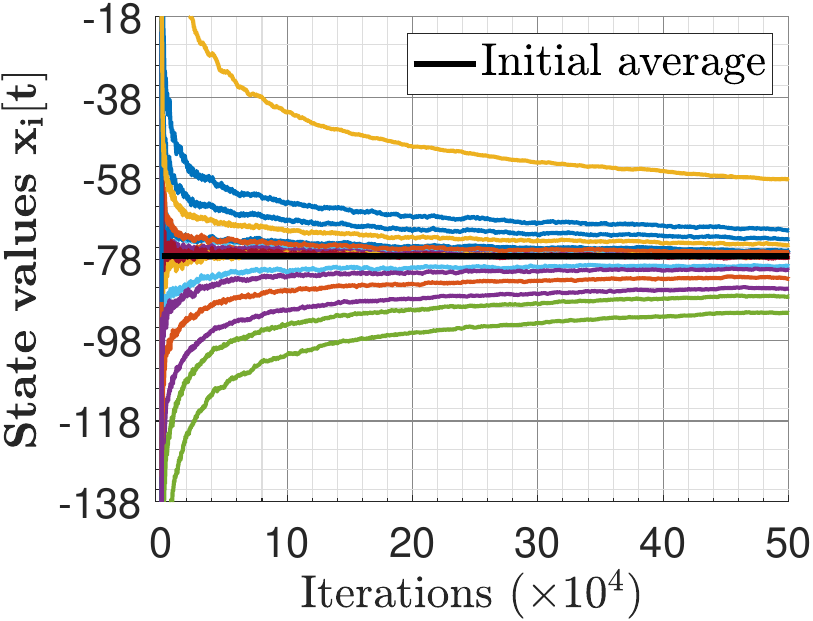}
\label{Fig.3:sub4}
}
\caption{The convergence dynamics of $x_{n}[t]$ for $18$ nodes in $5 \times 10^5$ iterations.
}
\label{Fig.3}
\end{figure}

\begin{figure}[h!]
\centering
\subfloat[]{
\includegraphics[width=0.22\textwidth]{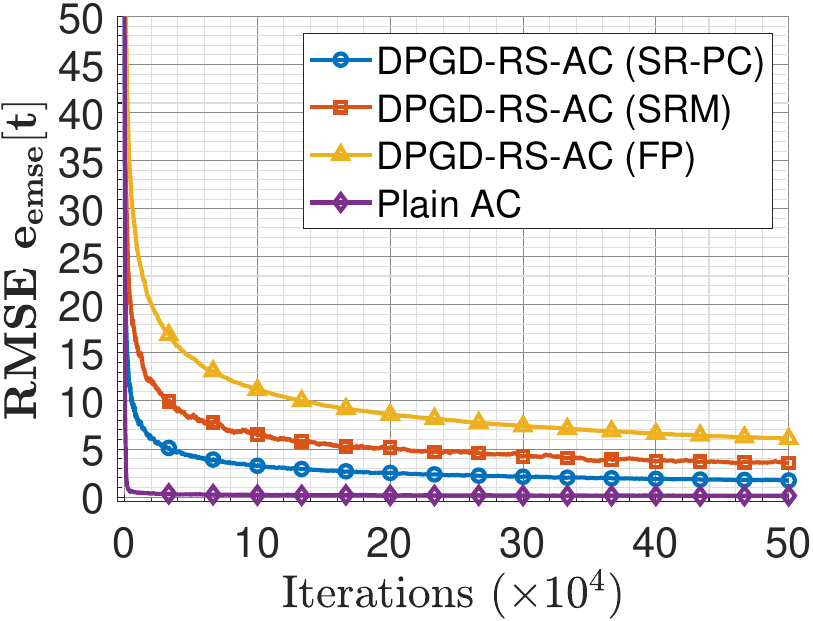}
\label{Fig.4:sub1}
}
\hspace{0pt}
\subfloat[]{
\includegraphics[width=0.22\textwidth]{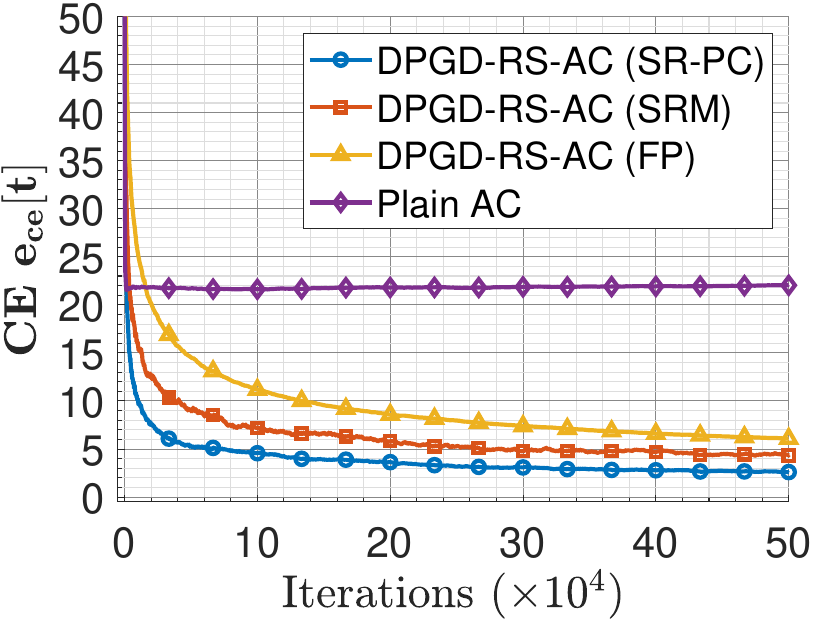}
\label{Fig.4:sub2}
}
\caption{Performance comparison. (a) Root mean square error (RMSE). (b) Consensus error (CE).}
\label{Fig.4}
\end{figure}

The convergence dynamics of $x_{n}[t]$ for all $n \in \mathcal{N}$ is given in Fig. \ref{Fig.3}, where different schemes start from identical initial conditions, share the same network topology, and experience the same small-scale fading $\mathbf{h}_{n,\,m}[t]$ in each time index $t$. 
The convergence performance of different consensus protocols and power-control schemes are compared in Fig. \ref{Fig.4}. For performance evaluation, we consider two metrics: consensus error (CE) $e_{\mathrm{ce}}[t]$ and root mean square error (RMSE) $e_{\mathrm{rmse}}[t]$, which are defined as
\begin{align}
\label{Eq.performance evaluation}	
&e_{\mathrm{ce}}[t] = \sqrt{\sum\limits_{n=1}^{N} (x_{n}[t] - x^{*})^{2}/N};  \\
&e_{\mathrm{rmse}}[t] =  \sqrt{\sum\limits_{n=1}^{N} \left(x_{n}[t] - \sum\limits_{m=1}^{N} x_{m}[t] /N \right)^{2}/N}.
\end{align}
Note that $e_{\mathrm{ce}}[t]$ serve as a Monte Carlo estimate of $E_{\mathrm{ce}}[t]$ defined in \eqref{Eq. root mean-square CE}, and quantifies the bias of the consensus value from the initial average $x^*$; while $e_{\mathrm{rmse}}[t]$ reflects the disagreement among agents. 

We implement a two-stage performance evaluation which isolates the improvements due to the proposed DPGD-RS-AC algorithm from those related to the power-control and receive-scaling schemes. First, we compare the proposed DPGD-RS-AC algorithm \eqref{Eq.UpdatedRule4} with the PAC protocol \eqref{Eq.benchmark} in \cite{10607888Yang}.
To ensure a fair comparison, we enforce the same average power consumption for both the algorithms; however, the power factors of DPGD-RS-AC is given by SR-PC, whereas PAC assigns identical power factor to each agent, as described in \cite{10607888Yang}. Subsequently, we evaluate different power-control and receive-scaling schemes using in the DPGD-RS-AC algorithm, with $\alpha_{\max} = \gamma_{\max} = 50$. In particular, the proposed SR-PC scheme is compared with the two benchmark methods: FP and SRM described in Section~\ref{sec:alternative}.

\subsection{Comparison Between Different Consensus Algorithms}
In this section, we compare the proposed DPGD-RS-AC algorithm in \eqref{Eq.UpdatedRule4} with the PAC protocol \eqref{Eq.benchmark} in \cite{10607888Yang}. As shown in Fig. \ref{Fig.3}, the limiting consensus value of DPGD-RS-AC nearly coincides with the initial average $x^*$, without any observable steady-state bias. In contrast, although the PAC protocol converges faster, it settles at a certain value that deviates from $x^*$. This trend is also confirmed in Fig. \ref{Fig.4}: The CE of DPGD-RS-AC continues to decrease toward $0$, whereas that of PAC plateaus at a non-zero value. 

The bias in PAC mainly stems from the existence of non-coherent interference ($c_{2,n}[t]$) and aggregated noise ($c_{3,n}[t]$), which prevent the mixing matrix $\mathbf{W}[t]$ from being row and column stochastic. Consequently, the average of the agents' initial states is not preserved in the network over iterations. Thanks to the decreasing stepsize $\zeta[t]$, the contributions of $c_{2,n}[t]$ and $c_{3,n}[t]$ finally vanish as $t \rightarrow \infty$, and agents can ultimately reach a consensus, although not on the exact $x^*$.

This result indicates that the proposed DPGD-RS-AC algorithm can achieve MSAC and resolve the issue of consensus bias in the PAC protocol \cite{10607888Yang}, highlighting its robustness. Incorporating the gradient term $\mathbf{d_q}[t]$ contributes to this robustness, since it acts as a regularization against interference and noise.  

\subsection{Impact of Different Power Control Schemes}
In this section, we compare the performance of three different power-control and signal-scaling schemes: SR-PC, FP and SRM, where the SR-PC scheme is our preference, while FP and SRM are included as benchmarks. Fig. \ref{Fig.3} provides an intuitive comparison of the convergence dynamics under the three schemes. In particular, SR-PC achieves the fastest convergence and the smallest convergence error, while exhibiting smooth convergence dynamics. SRM remains attractive since it is easier to obtain and offers a reasonable, albeit suboptimal, convergence rate. 
However, as shown in Fig. \ref{Fig.3:sub2}, SRM exhibits more pronounced fluctuations, consistent with its focus on minimizing the spectral radius without explicitly accounting for aggregation errors induced by non-coherent interference and receiver noise. Finally, FP exhibits the worst performance.

Consistent with these observations, Fig. \ref{Fig.4} further corroborates the same conclusion in terms of quantitative metrics: for both RMSE and CE, SR-PC shows the fastest decrease toward $0$. These results validate the effectiveness of our power-control and receive-scaling design in accelerating average-consensus convergence in wireless multi-agent systems employing NCOTA aggregation.

\section{Conclusion}
We investigated NCOTA aggregation in wireless multi-agent systems aiming at achieving distributed average consensus. Two major enhancements were developed to improve both the convergence speed and accuracy of existing NCOTA consensus algorithms:
\begin{itemize}
	\item We adopted a decentralized projected gradient method with row-stochastic mixing to design the encoding and decoding functions, which guarantees mean-square convergence to the exact average consensus value in the presence of persistent aggregation error.
	\item We developed a power control optimization framework that jointly optimizes transmit power allocation and receive scaling coefficients to accelerate convergence. The resulting scheme significantly outperforms full-power transmission and spectral-radius-minimizing benchmarks, demonstrating the benefits of enhancing effective network connectivity through physical-layer signal scaling.
\end{itemize}
In summary, our work provides a highly scalable, resource-efficient, and noise-resilient communication framework for multi-agent consensus and cooperation over wireless networks.

\appendices
\renewcommand\thesubsection{\thesection.\Roman{subsection}} 

\setcounter{subsection}{0} %
\def\thesubsectiondis{Appendix B.\Roman{subsection}:} 
\section*{Appendix A: Proof of Lemma \ref{Lemma.4}}
For a random vector $\mathbf{a}$, we define its variance as $\operatorname{var}\left(\mathbf{a}\right)  \triangleq \mathbb{E}\left[\| \mathbf{a} - \mathbb{E}\left[\mathbf{a}\right] \|^2\right]$, and its standard deviation as $\operatorname{sdv}\left(\mathbf{a}\right) \triangleq \sqrt{\operatorname{var}\left(\mathbf{a}\right)}$. 
Without loss of generality, we present the proof for the iteration $t = t_0$, while the proof can be readily generalized to any $t \in \mathbb{N}_0$. For simplicity, we denote $\operatorname{sdv}\left(\gamma_n c_n[t_0] \,\vert\, \mathcal{F}_{t_0}\right)$ as $\operatorname{sdv}\left(\gamma_n c_n\right)$ and $\mathbb{E}\left[\| \bm\epsilon[t_0]\|^2 \vert \mathcal{F}_{t_0}\right]$ as $\mathbb{E}\left[\| \bm\epsilon\|^2\right]$. To start with, we show that:
\begin{align}
\label{Eq.sum_sdv^2}
& \sum_{n=1}^{N} \operatorname{sdv}^2 \left(\gamma_n c_n\right) = \sum_{n=1}^{N} \mathbb{E}\left[\left( \gamma_n c_n - \mathbb{E}\left[\gamma_n c_n\right] \right)^2 \right] \nonumber \\
= & \sum_{n=1}^{N} \mathbb{E}\left[ \lvert \epsilon_n \rvert^2\right] = \mathbb{E}\left[\| \bm\epsilon\|^2\right].
\end{align}

In order to bound $\mathbb{E}\left[\| \bm\epsilon\|^2\right]$, we first derive the upper bound of $\operatorname{sdv}\left(\gamma_n c_n\right)$ for all $n \in \mathcal{N}$:
\begin{align}
\label{Eq.sdv_reveiced_symbol}
\operatorname{sdv}\left(\gamma_n c_n\right) 
&= \gamma_n \sqrt{ \mathbb{E}\left[\left( c_n - \mathbb{E}\left[ c_n\right] \right)^2 \right]} \\
& \leq  \gamma_n \sum_{i=1}^{2} \Big(\sqrt{\mathbb{E}\left[ \left(\lvert r_{n,i} \rvert^2 \!-\! \mathbb{E}\left[ \lvert r_{n,i}\rvert^2  \right] \right)^2 \right]}  \lvert z_i \!-\! x_n\rvert\Big), \nonumber
\end{align}
where the inequality holds due to the Minkowski inequality \cite[Lemma $14.10$]{FlorescuTudor2013}: 
$$\sqrt{\mathbb{E}\left[\big\| \sum_{n=1}^{N} \mathbf{a}_n \big\|^2\right]} \leq \sum_{n=1}^{N} \sqrt{\mathbb{E}\left[\|\mathbf{a}_n \|^2\right]}$$
for any random vector $\mathbf{a}_n$ where $n \in \mathcal{N}$.
For $\lvert z_i - x_n\rvert$, we show that:  
\begin{align}
&\lvert z_i - x_n \rvert \leq \sum_{m=1}^{2} p_{n,m} \lvert z_i - z_m\rvert \nonumber \\ 
\leq & \max_{m\in\{1,\,2\}} \lvert z_i - z_m\rvert_2 \leq \max_{z, z^{\prime} \in \mathcal{Z}} \lvert z - z^{\prime}\rvert = 2x_{\max}.
\end{align}
In addition, we give the close-form expression of $$\operatorname{sdv}\left( \lvert r_{n,i}\rvert^2\right) = \sqrt{\mathbb{E}\left[ \left(\lvert r_{n,i} \rvert^2 - \mathbb{E}\left[ \lvert r_{n,i}\rvert^2  \right] \right)^2 \right]}.$$
According to \eqref{Eq.received_symbol}, we know that $r_{n,i} \sim \mathcal{CN}\left(0,\,\Gamma_{n,i} + \sigma^2\right)$, where $\Gamma_{n,i} = \sum_{m=1, m\neq n}^{N} \beta_{n,m}\alpha_m p_{m,i}$. Therefore, $\lvert r_{n,i}\rvert^2$ follows an exponential distribution with parameter $\lambda = 1/\left(\Gamma_{n,i} + \sigma^2\right)$, i.e., $\lvert r_{n,i}\rvert^2 \sim \operatorname{Exp}\left(1/\left(\Gamma_{n,i}+\sigma^2\right)\right)$. Consequently, $\operatorname{sdv}\left( \lvert r_{n,i}\rvert^2\right) = \Gamma_{n,i} + \sigma^2$. Plugging the expression of $\operatorname{sdv}\left( \lvert r_{n,i}\rvert^2\right)$ in \eqref{Eq.sum_sdv^2} and \eqref{Eq.sdv_reveiced_symbol} yields:
\begin{align}
\label{Eq.error_bound}
&\frac{1}{N}\mathbb{E}\left[\| \bm\epsilon\|^2 \right] = \frac{1}{N}\sum_{n=1}^{N}\left(\operatorname{sdv}\left(\gamma_nc_n\right)\right)^2 \\
\leq &\frac{1}{N}\sum_{n=1}^{N}\left\{2 x_{\max} \gamma_n\left(\sum\limits_{\substack{m=1 \\ m\neq n}}^{N} \beta_{n,m}\alpha_{m} + 2\sigma^2 \right) \right\}^2 = \Sigma. \nonumber
\end{align}

\bibliographystyle{IEEEtran}
\bibliography{ref}

@INPROCEEDINGS{icc25,
	author={Deng, Yuhang and Chen, Zheng and Larsson, Erik G.},
	booktitle={Proc. IEEE Int. Conf. Commun. (ICC)}, 
	title={Robust and Efficient Average Consensus with Non-Coherent Over-the-Air Aggregation}, 
	year={2025},
	volume={},
	number={},
	pages={3455-3460},
	doi={10.1109/ICC52391.2025.11161728}}

@ARTICLE{1638541Boyd,
  author={Boyd, S. and Ghosh, A. and Prabhakar, B. and Shah, D.},
  journal={IEEE Trans. Inf. Theory}, 
  title={Randomized gossip algorithms}, 
  year={2006},
  volume={52},
  number={6},
  pages={2508-2530},
  doi={10.1109/TIT.2006.874516}}

@ARTICLE{4237149Mehyar,
  author={Mehyar, Mortada and Spanos, Demetri and Pongsajapan, John and Low, Steven H. and Murray, Richard M.},
  journal={IEEE/ACM Trans. Netw.}, 
  title={Asynchronous Distributed Averaging on Communication Networks}, 
  year={2007},
  volume={15},
  number={3},
  pages={512-520},
  doi={10.1109/TNET.2007.893226}}

@ARTICLE{4118472ReZa,
	author={Olfati-Saber, Reza and Fax, J. Alex and Murray, Richard M},
	journal={Proc. IEEE}, 
	title={Consensus and Cooperation in Networked Multi-Agent Systems}, 
	year={2007},
	volume={95},
	number={1},
	pages={215-233},
	doi={10.1109/JPROC.2006.887293}}

@ARTICLE{4140748RenWei,
	author={Ren, Wei and Beard, Randal W. and Atkins, Ella M.},
	journal={IEEE Control Syst. Mag.}, 
	title={Information consensus in multivehicle cooperative control}, 
	year={2007},
	volume={27},
	number={2},
	pages={71-82},
	doi={10.1109/MCS.2007.338264}}

@INPROCEEDINGS{1440896XiaoL,
	author={Xiao, L. and Boyd, S. and Lall, S.},
	booktitle={Proc. 4th Int. Symp. Inf. Process. Sensor Netw. (IPSN)}, 
	title={A scheme for robust distributed sensor fusion based on average consensus}, 
	year={2005},
	volume={},
	number={},
	pages={63-70},
	doi={10.1109/IPSN.2005.1440896}}

@book{ren2010distributed,
	title={Distributed Coordination of Multi-Agent Networks: Emergent Problems, Models, and Issues},
	author={Ren, Wei and Cao, Yongcan},
	year={2011},
	publisher={Springer},
    address={London, U.K.}
}

@article{degroot1974reaching,
  title={Reaching a consensus},
  author={DeGroot, Morris H},
  journal={J. Amer. Statist. Assoc.},
  volume={69},
  number={345},
  pages={118--121},
  year={1974},
  doi={10.1080/01621459.1974.10480137},
  publisher={Taylor \& Francis}
}

@ARTICLE{1333204Saber,
  author={Olfati-Saber, R. and Murray, R.M.},
  journal={IEEE Trans. Autom. Control}, 
  title={Consensus problems in networks of agents with switching topology and time-delays}, 
  year={2004},
  volume={49},
  number={9},
  pages={1520-1533},
  doi={10.1109/TAC.2004.834113}}

@article{200465Lin,
	title = {Fast linear iterations for distributed averaging},
	journal = {Syst. Control Lett.},
	volume = {53},
	number = {1},
	pages = {65-78},
	year = {2004},
	issn = {0167-6911},
	doi = {10.1016/j.sysconle.2004.02.022},
	author = {Lin Xiao and Stephen Boyd},
}

@article{olshevsky2009convergence,
	title={Convergence speed in distributed consensus and averaging},
	author={Olshevsky, Alex and Tsitsiklis, John N},
	journal={SIAM J. Control Optim.},
	volume={48},
	number={1},
	pages={33--55},
	year={2009},
    doi = {10.1137/060678324},
	publisher={SIAM}
}

@ARTICLE{4663899Kar,
	author={Kar, Soummya and Moura, Jos{\'e} M. F.},
	journal={IEEE Trans. Signal Process.}, 
	title={Distributed Consensus Algorithms in Sensor Networks With Imperfect Communication: Link Failures and Channel Noise}, 
	year={2009},
	volume={57},
	number={1},
	pages={355-369},
	doi={10.1109/TSP.2008.2007111}}

@article{06067359XHuang,
	author = {Huang, Minyi and Manton, Jonathan H.},
	title = {Coordination and Consensus of Networked Agents with Noisy Measurements: Stochastic Algorithms and Asymptotic Behavior},
	journal = {SIAM J. Control Optim.},
	volume = {48},
	number = {1},
	pages = {134-161},
	year = {2009},
	doi = {10.1137/06067359X}
}

@inproceedings{carli2007average,
	title={Average consensus on networks with transmission noise or quantization},
	author={Carli, Ruggero and Fagnani, Fabio and Frasca, Paolo and Taylor, Tom and Zampieri, Sandro},
	booktitle={Proc. Eur. Control Conf. (ECC)},
	pages={1852--1857},
    doi = {10.23919/ECC.2007.7068829},
	year={2007}
}

@ARTICLE{5325701Kar,
	author={Kar, Soummya and Moura, Jos{\'e} M. F.},
	journal={IEEE Trans. Signal Process.}, 
	title={Distributed Consensus Algorithms in Sensor Networks: Quantized Data and Random Link Failures}, 
	year={2010},
	volume={58},
	number={3},
	pages={1383-1400},
	doi={10.1109/TSP.2009.2036046}}

@ARTICLE{5929538Lavaei,
	author={Lavaei, Javad and Murray, Richard M.},
	journal={IEEE Trans. Autom. Control}, 
	title={Quantized Consensus by Means of Gossip Algorithm}, 
	year={2012},
	volume={57},
	number={1},
	pages={19-32},
	doi={10.1109/TAC.2011.2160593}}

@article{HUANG20101571,
	title = {Stochastic consensus over noisy networks with {Markovian} and arbitrary switches},
	journal = {Automatica},
	volume = {46},
	number = {10},
	pages = {1571-1583},
	year = {2010},
	issn = {0005-1098},
	doi = {10.1016/j.automatica.2010.06.016},
	author = {Minyi Huang and Subhrakanti Dey and Girish N. Nair and Jonathan H. Manton}
}

@article{fagnani2009average,
	title={Average consensus with packet drop communication},
	author={Fagnani, Fabio and Zampieri, Sandro},
	journal={SIAM J. Control Optim.},
	volume={48},
	number={1},
	pages={102--133},
	year={2009},
    doi={10.1137/060676866},
	publisher={SIAM}
}

@INPROCEEDINGS{5499113Shen,
	author={Shen Wang and Cuijuan An and Xiao-Xiang Sun and Xin Du},
	booktitle={Proc. Chinese Control Decis. Conf. (CCDC)}, 
	title={Average consensus over communication channels with uniform packet losses}, 
	year={2010},
	volume={},
	number={},
	pages={114-119},
	doi={10.1109/CCDC.2010.5499113}}

@article{10092857sahin,
	title={A survey on over-the-air computation},
	author={{\c{S}}ahin, Alphan and others},
	journal={IEEE Commun. Surveys Tuts.},
	volume={25},
	number={3},
	pages={1877--1908},
	year={2023},
	publisher={IEEE}
}

@ARTICLE{10155550Chen,
	author={Chen, Zheng and Larsson, Erik G. and Fischione, Carlo and Johansson, Mikael and Malitsky, Yura},
	journal={IEEE Netw.}, 
	title={Over-the-Air Computation for Distributed Systems: Something Old and Something New}, 
	year={2023},
	volume={37},
	number={5},
	pages={240-246},
	doi={10.1109/MNET.126.2200205}}

@ARTICLE{9958941Lin,
	author={Lin, Zhenyi and Gong, Yi and Huang, Kaibin},
	journal={IEEE J. Sel. Areas Commun.}, 
	title={Distributed Over-the-Air Computing for Fast Distributed Optimization: Beamforming Design and Convergence Analysis}, 
	year={2023},
	volume={41},
	number={1},
	pages={274-287},
	doi={10.1109/JSAC.2022.3223661}}

@ARTICLE{10607888Yang,
	author={Yang, Huiwen and Chen, Xiaomeng and Huang, Lingying and Dey, Subhrakanti and Shi, Ling},
	journal={IEEE Trans. Control Netw. Syst.}, 
	title={Distributed Average Consensus via Noisy and NonCoherent Over-the-Air Aggregation}, 
	year={2025},
	volume={12},
	number={1},
	pages={64--73},
	doi={10.1109/TCNS.2024.3432827}}

@ARTICLE{michelusi2024ota,
	author={Michelusi, Nicol{\o}`},
	journal={IEEE Trans. Signal Process.}, 
	title={Non-Coherent Over-the-Air Decentralized Gradient Descent}, 
	year={2024},
	volume={72},
	number={},
	pages={4618-4634},
	doi={10.1109/TSP.2024.3460690}}

@ARTICLE{4749425Nedic,
	author={Nedic, Angelia and Ozdaglar, Asuman},
	journal={IEEE Trans. Autom. Control}, 
	title={Distributed Subgradient Methods for Multi-Agent Optimization}, 
	year={2009},
	volume={54},
	number={1},
	pages={48-61},
	doi={10.1109/TAC.2008.2009515}}

@book{FlorescuTudor2013,
	author    = {Florescu, Ionut and Tudor, Caius Adrian},
	title     = {Handbook of Probability},
	publisher = {Wiley},
    address = {Hoboken, NJ, USA},
	year      = {2013}
}

@book{nocedal2006numerical,
	title     = {Numerical Optimization},
	author    = {Nocedal, Jorge and Wright, Stephen J.},
	edition   = {2nd},
	year      = {2006},
	publisher = {Springer},
	address   = {New York, NY, USA},
	isbn      = {978-0-387-30303-1}
}

@article{lobel2011distributed,
	title={Distributed multi-agent optimization with state-dependent communication},
	author={Lobel, Ilan and Ozdaglar, Asuman and Feijer, Diego},
	journal={Math. Program.},
	volume={129},
	number={2},
	pages={255--284},
	year={2011},
    doi={10.1007/s10107-011-0467-x},
	publisher={Springer}
}

@ARTICLE{5411807LiTao,
	author={Li, Tao and Zhang, Ji-Feng},
	journal={IEEE Trans. Autom. Control}, 
	title={Consensus Conditions of Multi-Agent Systems With Time-Varying Topologies and Stochastic Communication Noises}, 
	year={2010},
	volume={55},
	number={9},
	pages={2043-2057},
	doi={10.1109/TAC.2010.2042982}}

@INPROCEEDINGS{9322286Ozfatura,
	author={Ozfatura, E. and Rini, Stefano and G{\"u}nd{\"u}z, D.},
	booktitle={Proc. IEEE Global Commun. Conf. (GLOBECOM)}, 
	title={Decentralized {SGD} with Over-the-Air Computation}, 
	year={2020},
	volume={},
	number={},
	pages={1-6},
	doi={10.1109/GLOBECOM42002.2020.9322286}}

@ARTICLE{10154060Nikos,
	author={Mitsiou, Nikos A. and others},
	journal={IEEE Trans. Commun.}, 
	title={Accelerating Distributed Optimization via Over-the-Air Computing}, 
	year={2023},
	volume={71},
	number={9},
	pages={5565-5579},
	doi={10.1109/TCOMM.2023.3286915}}

@misc{deng2026exactaverageconsensusnoisy,
	title={Exact Average Consensus under Noisy Communication Links: A Decentralized Gradient Perspective}, 
	author={Yuhang Deng and Zheng Chen and Erik G. Larsson},
	year={2026},
	eprint={2609.28082},
	archivePrefix={arXiv},
	primaryClass={eess.SY},
    note={arXiv:2609.28082},
	url={https://arxiv.org/abs/2609.28082}, 
}

@article{MAI201994,
title = {Distributed optimization over directed graphs with row stochasticity and constraint regularity},
journal = {Automatica},
volume = {102},
pages = {94-104},
year = {2019},
issn = {0005-1098},
doi = {10.1016/j.automatica.2018.07.020},
author = {Van Sy Mai and Eyad H. Abed}
}

@ARTICLE{11543335Dahl,
  author={Dahl, Martin and Chen, Zheng and Larsson, Erik G.},
  journal={IEEE Trans. Commun.}, 
  title={A Unified Framework for Unbiased Non-Coherent Over-the-Air Computation}, 
  year={2026},
  volume={74},
  number={},
  pages={11124-11139},
  doi={10.1109/TCOMM.2026.3698903}}

@INPROCEEDINGS{11587673Michelusi,
  author={Michelusi, Nicol{\o}`},
  booktitle={Proc. IEEE Int. Conf. Commun. (ICC)}, 
  title={Interference-Robust Non-Coherent Over-the-Air Computation for Decentralized Optimization}, 
  year={2026},
  volume={},
  number={},
  pages={1-6},
  doi={10.1109/ICC59461.2026.11587673}}

@misc{2026Bozorgasl,
      title={Resource-Element Energy Difference for Noncoherent Over-the-Air Federated Learning}, 
      author={Hao Chen and Zavareh Bozorgasl},
      year={2026},
      eprint={2605.07263},
      archivePrefix={arXiv},
      primaryClass={eess.SP},
      note={arXiv:2605.07263},
      url={https://arxiv.org/abs/2605.07263}, 
}

@ARTICLE{11250607Chen,
  author={Chen, Xiaomeng and Yang, Huiwen and Dey, Subhrakanti and Shi, Ling},
  journal={IEEE Trans. Signal Process.}, 
  title={Non-Coherent Over-the-Air Decentralized Method for Non-Cooperative Games in Multi-Agent Systems}, 
  year={2025},
  volume={73},
  number={},
  pages={4684-4699},
  doi={10.1109/TSP.2025.3632064}}

@misc{Haifeng2026,
      title={Non-Coherent Over-the-Air Federated Learning: Protocol, Convergence, and Device Scheduling}, 
      author={Haifeng Wen and Nicolò Michelusi and Osvaldo Simeone and Yang Yang and Hong Xing},
      year={2026},
      eprint={2609.08312},
      archivePrefix={arXiv},
      primaryClass={cs.IT},
      note={arXiv:2609.08312},
      url={https://arxiv.org/abs/2609.08312}, 
}

@book{goldsmith2005wireless,
	title={Wireless Communications},
	author={Goldsmith, Andrea},
	year={2005},
    address={Cambridge, U.K.},
	publisher={Cambridge Univ. Press}
}
\end{document}